\documentclass[fleqn,usenatbib,referee]{mnras}

\usepackage{hyperref}
\usepackage[T1]{fontenc}
\usepackage{ae,aecompl}
\usepackage{times}
\usepackage{graphicx}	% Including figure files
\usepackage{amsmath}	% Advanced maths commands
\usepackage{amssymb}	% Extra maths symbols
\usepackage{enumitem}
\usepackage{longtable}
\usepackage{lscape}
\usepackage{pdflscape}
\usepackage{tabularx} % in the preamble
\usepackage[a4paper]{geometry}
\title[Physical properties of star clusters in the outer LMC as observed by the Dark Energy Survey]{Physical properties of star clusters in the outer LMC as observed by the DES}
\author[Pieres,~A. et al.]{A. Pieres\thanks{E-mail: adriano.pieres@linea.gov.br}$^{1,2}$, B. X. Santiago$^{1,2}$, E. Balbinot$^3$, E. Luque$^{1,2}$, A. B. A. Queiroz$^{1,2}$, 
\newauthor
L. N. da Costa$^{2,4}$, M. A. G. Maia$^{2,4}$, A. Drlica-Wagner$^{5}$, A.~Roodman$^{6,7}$, 
\newauthor
C.~Abbott$^{8}$, S.~Allam$^{5}$, A.~Benoit-L{\'e}vy$^{9}$, E.~Bertin$^{10,11}$, D.~Brooks$^{9}$,
\newauthor
E.~Buckley-Geer$^{5}$, D.~L.~Burke$^{6,7}$, A.~Carnero~Rosell$^{2,4}$, M.~Carrasco~Kind$^{12,13}$,
\newauthor
J.~Carretero$^{14,15}$, C.~E.~Cunha$^{6}$, S.~Desai$^{16,17}$, H.~T.~Diehl$^{5}$, T.~F.~Eifler$^{18,19}$,
\newauthor
D.~A.~Finley$^{5}$, B.~Flaugher$^{5}$, P.~Fosalba$^{14}$, J.~Frieman$^{5,6}$, D.~W.~Gerdes$^{20}$,
\newauthor
D.~Gruen$^{21,22}$, R. A. Gruendl$^{23,24}$, G.~Gutierrez$^{5}$, K.~Honscheid$^{25,26}$,
\newauthor
D.~J.~James$^{8}$, K.~Kuehn$^{27}$, N.~Kuropatkin$^{5}$, O.~Lahav$^{9}$, T.~S.~Li$^{28}$, J.~L.~Marshall$^{28}$, 
\newauthor
P. Martini$^{25,29}$, C.~J.~Miller$^{20,30}$, R.~Miquel$^{15,31}$, R.~C.~Nichol$^{32}$, B.~Nord$^{5}$,
\newauthor
R.~Ogando$^{2,4}$, A.~A.~Plazas$^{19}$, A.~K.~Romer$^{33}$, E.~Sanchez$^{34}$, V.~Scarpine$^{5}$, 
\newauthor
M.~Schubnell$^{20}$, I.~Sevilla-Noarbe$^{12,34}$, R.~C.~Smith$^{6}$, M.~Soares-Santos$^{5}$, 
\newauthor
F.~Sobreira$^{2,5}$, E.~Suchyta$^{25,26}$, M.~E.~C.~Swanson$^{13}$, G.~Tarle$^{20}$, J.~Thaler$^{35}$, 
\newauthor
D.~Thomas$^{32}$, D.~Tucker$^{5}$, A. R. Walker$^{8}$ \\
Affiliations are listed after the references}

\date{Last updated 2015 Nov 06; in original form 2015 November 06}

\pubyear{2015}
\begin{document}
\label{firstpage}
\pagerange{\pageref{firstpage}--\pageref{lastpage}}

%\pagerange{\pageref{firstpage}--\pageref{lastpage}} \pubyear{XXXX}
\maketitle

\begin{abstract}
The Large Magellanic Cloud (LMC) harbors a rich and diverse system of star clusters, whose ages, chemical abundances, and positions provide information about the LMC history of star formation. We use Science Verification imaging data from the Dark Energy Survey to increase the census of known star clusters in the outer LMC and to derive physical parameters for a large sample of such objects using a spatially and photometrically homogeneous data set.
Our sample contains 255 visually identified cluster candidates, of which 109 were not listed in any previous catalog. We quantify the crowding effect for the stellar sample produced by the DES Data Management pipeline and conclude that the stellar completeness is $<$ 10\% inside typical LMC cluster cores. We therefore develop a pipeline to sample and measure stellar magnitudes and positions around the cluster candidates using \textsc{daophot}. We also implement a maximum-likelihood method to fit individual density profiles and colour-magnitude diagrams. For 117 (from a total of 255) of the cluster candidates (28 uncatalogued clusters), we obtain reliable ages, metallicities, distance moduli and structural parameters, confirming their nature as physical systems. The distribution of cluster metallicities shows a radial dependence, with no clusters more metal-rich than $[Fe/H] \simeq -0.7$ beyond 8 kpc from the LMC center. The age distribution has two peaks at $\simeq$ 1.2 Gyr and $\simeq$ 2.7 Gyr.
\end{abstract}


\section{Introduction}
\label{introd}
\smallskip
% Two paragraphs about LMC and its resovible population
The LMC is a nearby dynamically active satellite galaxy, exhibiting multiple epochs of star formation, while also suffering from tidal interactions with the Small Magellanic Cloud (SMC) and the Milky Way (MW). Given its proximity, stellar populations in the LMC are easily resolved in deep surveys, allowing us to obtain information such as ages, chemical abundances, kinematics and distances to individual stars and star clusters. Thus, the LMC represents an excellent local laboratory to study the effects of gravitational forces on the evolution of a satellite galaxy, including its star formation history (SFH) and age-metallicity relation (AMR).

A wealth of data describing the structure and stellar populations of the LMC has been accumulated over decades of research. The LMC is known to have a stellar disk inclined relative to the line of sight towards its center by $i = 36-38^{\circ}$ with a position angle of $\theta = 130-145^{\circ}$. This disk also seems to have a warp and to be flared \citep{1986MNRAS.218..223C, 2002AJ....124.2045O, 2010A&A...520A..24S, 2015MNRAS.449.1129B}. A large number of studies have tried to reconstruct its SFH and/or AMR, often in connection to the SMC \citep{2008AJ....135..836C, 2009AJ....138.1243H, 2010A&A...517A..50G, 2011AJ....142...61C, 2011A&A...535A.115I, 2012A&A...537A.106R, 2013AJ....145...17P, 2013MNRAS.431..364W, 2014MNRAS.438.1067M}. 

Regarding the stellar populations and their variation as a function of position, many studies attempt to reconstruct the SFH and/or AMR of the LMC, based on field stars~\citep{2014MNRAS.438.1067M, 2011AJ....142...61C, 2008AJ....135..836C}, clusters in the relatively central region ($<$ 10 kpc)~\citep{2010A&A...517A..50G, 2007IAUS..235...92G, 2007A&A...462..139K, 1997AJ....114.1920G, 1993A&A...269..107K, 1991AJ....101..515O} or both~\citep{2013AJ....145...17P}. The results have been inconsistent, as no single SFH and AMR applies to the entire LMC body or to clusters and field stars alike \citep{2013AJ....145...17P, 2011AJ....142...61C}. In a recent paper, \cite{2013AJ....145...17P} analyse 5.5 million LMC field stars and present age and metallicity trends with distance from the LMC center (out to 8 kpc) and an AMR. Their results are more consistent with outside-in star formation and chemical enrichment. They also find larger spreads in age and metallicity in the outer regions and no clear age gap in the field star formation. \cite{2011AJ....142...61C} investigate fields stars farther out, from 5.2 to 9.2 kpc from the LMC center, and find age and metallicity gradients only for the youngest and the most metal rich stars. For the star clusters,~\cite{2010A&A...517A..50G} fit Padova \citep{2002A&A...391..195G} and Geneva~\citep{2001A&A...366..538L} isochrones to a sample of 1193 young LMC clusters within 4 deg of its center and find two periods of enhanced cluster formation, at 125 Myr and 800 Myr. They argue that these peaks in the cluster formation rate may be connected with the last encounter of the LMC and SMC.

In contrast with our knowledge of the inner structure and stellar populations of the LMC, much less is known about the periphery at distances $>$ 10 kpc. Covering the extended outer LMC regions requires a large-area, photometric, and homogeneous sample. This has recently been provided by the early data taken as part of the Dark Energy Survey (DES) Science Verification (SV). In this work we aim to probe clusters in the outer LMC field in DES-SV footprint, using a homogeneous data sample. %We implemented methods to fit density profiles and compare data to isochrones, as well. These methods are validated on simulations and applied to the SV data.

This paper is organized as follows: Sec.~\ref{dados} is a brief introduction presenting the data (the DES-SV data set, its reduction, and the resulting LMC clusters sample). Sec.~\ref{methods} describes the methods we applied to recover cluster structural parameters and cluster stellar populations. In Sec.~\ref{clussimul} we tested the limits of the methods, simulating clusters, recovery of parameters and determining uncertainties. Section~\ref{truedata} presents our results and compares to avaliable literature. Section~\ref{summary} provides a discussion and summary.

\section{Data}
\label{dados}

\subsection{DES and Science Verification}
\label{desdata}
% About DES and SV
The DES \citep{2005astro.ph.10346T} is a 5,000 deg$^2$ imaging survey in \emph{grizY} bands currently being carried out using DECam, a 3 deg$^2$ ($2.2\,^{\circ}$ diameter) wide-field mosaic camera on the CTIO Blanco 4\,m telescope \citep{2015arXiv150402900F}. DES will reach a characteristic photometric depth of $24^{th}$ magnitude (with S/N=~10 for $g$ band, point-like sources) and enable accurate photometry and morphology of objects ten times fainter than the Sloan Digital Sky Survey (SDSS).

The DECam images are reduced by the DES Data Management (DESDM) team, which has a pipeline to coadd and calibrate (astrometrically and photometrically) images and finally catalog and classify the objects in the images. The final coadd images are called \emph{tiles}, with a size of 0.75 degree $\times$ 0.75 degree ($10^4 \times 10^4$ pixels). More details can be found in \cite{2012SPIE.8451E..0DM} and \cite{2012ApJ...757...83D}.

A substantial challenge for deep ground-based surveys is the star/galaxy separation. DESDM performs photometric analysis with \textsc{sextractor}, a software package which identifies and selects sources above a threshold based on the image background. The blend parameter determines whether a group of neighboring pixels must be classified as a single source or multiple objects \citep{2011ASPC..442..435B}. The final catalog produces several magnitude measurements (\emph{mag\_auto} and \emph{mag\_psf}, for example) and \emph{spread\_model} as the main star/galaxy separator. This is based on the difference between the best fitting local point spread function (PSF) model and a slightly more extended model made from the same PSF convolved with a circular exponential disk model with scale-length equal to FWHM/16 (where FWHM is the full-width-half maximum of the PSF model).

The DES-SV data were taken from November 2012 to February 2013 for a total of 427 hours of observation. The data were intended to test DECam capabilities and the DESDM data pipeline (calibration and photometry, astrometry, image quality, pointing and guiding, operational readiness and pipelines for supernova). The largest contiguous area of the SV campaign was the South Pole Telescope East field (SPT-E), which is located immediately to the north of the LMC. This work is based on the coadd data products from the first release processing of the SV observations (SVA1), which includes the SPT-E region (157 deg$^2$) in 2537 exposures. %More details on this data set can be found in~\citep{2014AAS...22314106R}.

\subsection{LMC star clusters}
\label{lmcclus}

An initial sample of LMC star clusters in the SPT-E region was obtained by a visual inspection of the SV coadd images, on a tile by tile basis. We used the Tile Viewer tool available at the DES Science portal~\citep{2012ASPC..461..287B}, which is a web based facility. The list of SPT-E tiles was initially split among five of the authors with an overlap region among them. Almost all LMC clusters were found in the southern part of the field, which was inspected by two of us. The overlapping region included ten clusters found by one of the authors, eight of which comprised the sample found by the other. Not unexpectedly, the two extra objects were poor clusters. Despite the low numbers, this sample overlap suggests a uniform sampling among the richer systems analyzed in \S \ref{clusanalysis}.

After eliminating repeats from the individual searches, the original cluster list contained 294 candidates. We then matched our candidates with the \cite{Bica2008} LMC clusters catalog. Our list contains a total of 109 previously uncatalogued objects. Seven objects included in the \cite{Bica2008} catalog were not in our original list. By visually inspecting their images, one of these objects was clearly a galaxy and another was very close to a bright star, leaving a total of 299 cluster candidates in the SPT-E region.

\begin{figure*}
\begin{minipage}{130mm}
\centering
\includegraphics[width=130mm,clip=]{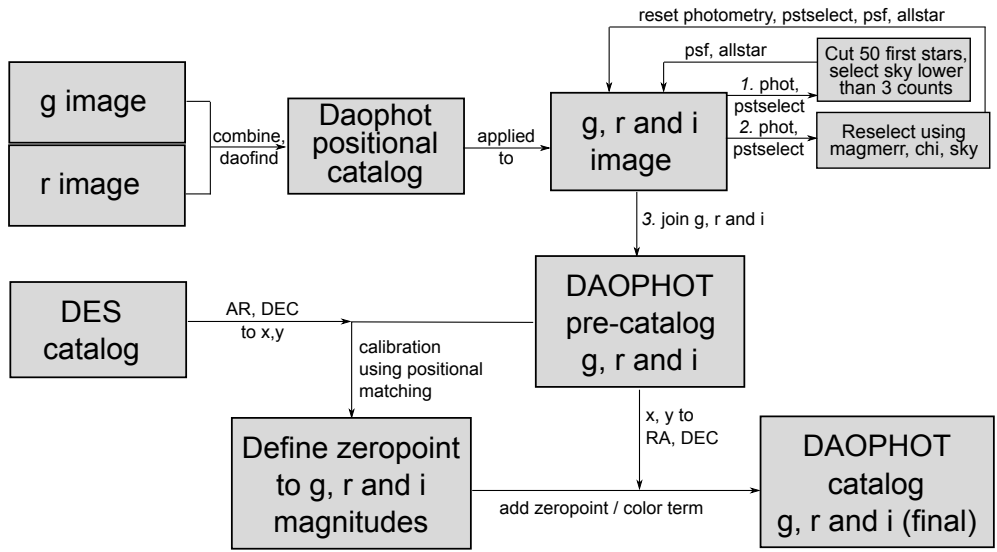}
\caption{Flow chart describing the \textsc{daophot} based reduction and photometry pipeline developed as part of this work. As final products, it yields a catalog using \emph{g, r} and \emph{i} magnitudes from coadd image cutouts. First, the \emph{g} and \emph{r} images are combined to generate a positional sources catalog (using \emph{daofind}). This catalog is a starting point to run \emph{phot} (for aperture measurements) and \emph{pstselect} (for producing a list of stars to be used in the PSF model) routines for each filter. We pick the 50 brightest sources and run \emph{psf} (that does the actual PSF modelling) and \emph{allstar} (PSF fitting to all detected sources) tasks. A cleaner set of psf stars is selected as those sources with the lowest values of \emph{chi, magerror, sky}, making up a new PSF list. We reset the list and rerun \emph{psf} and \emph{allstar}. The zeropoint for each filter are determined comparing magnitudes of SVA1 (DESDM) and \textsc{daophot} after a positional match to the DES catalog. Last, these calibration terms are added and the coordinates converted to $\alpha$ and $\delta$, generating the final composite \textsc{daophot} catalog.}
\label{flux}
\end{minipage}
\end{figure*}

The sample was further refined using the $g~vs.~g-r$ CMDs (from \textsc{daophot}, see \S \ref{daophotpipe}). 44 out of the 299 visually inspected candidates had CMDs that were not consistent with simple or composite stellar populations, usually containing a large fraction of faint red sources, more consistent with galaxy clusters. The filtered subsample has 255 candidates consistent with being star clusters.

\subsection{DES-SV stellar sample}
\label{complet}

Since the DES main goals are cosmological in nature, the science requirements for the data are related to detecting, measuring and characterizing galaxies, not stars. Therefore, we ran a complementary diagnostic of the DES-SV stellar catalog. Our main concern was to investigate the completeness of the SVA1 stellar catalog as a function of S/N and crowding, since this aspect is crucial to the analysis of LMC clusters presented here. For that purpose, we used routines from \textsc{iraf/daophot}, which is the benchmark program to detect and measure stellar data in crowded fields. Details about our \textsc{daophot} reduction and photometry pipeline are provided in \S \ref{daophotpipe}. A full description of stellar completeness in SVA1 catalogs can be found in a future technical paper (Pieres et al 2015, in prep.). Below we summarize our basic conclusions:

\begin{enumerate}[leftmargin=.75cm,labelsep=0.2cm,align=left,label=(\roman*)]

\item Colour-magnitude diagrams based on the stellar samples drawn from standard \textsc{daophot} selection, using PSF magnitude errors and sharpness, are similar to those from SVA1, based on \emph{spread\_model} cuts. In the SPT-E fields covering the outer LMC, the CMDs from both SVA1 and \textsc{daophot} clearly display the main features of the LMC field population, such as a MS ranging from 18 $\textless$ g $\textless$ 24, an old MSTO at g$\sim$22.5, g-r $\sim$0.2, a red giant branch and a red clump with g $\textless$ 22 and g-r $\textgreater$ 0.3.

\item In typical DES regions, well away from the LMC, the source density is about 1/10 of that found in regions where most star clusters are found. In this low-density regime, as well as inside poor LMC clusters, SVA1 completeness relative to \textsc{daophot} is close to 1 down to g $\cong$ 20. SVA1 catalogs tend to sample fewer point sources than \textsc{daophot} at fainter magnitudes, but the relative completeness varies depending on how we separate stars from galaxies in either case.

\item SVA1 stellar completeness is a strong function of source density, dropping abruptly from $\sim$ 0.3 to $\textless$ 0.1 for surface densities $\textgreater$ 260 stars/arcmin$^2$. In particular, the SVA1 stellar sample is very incomplete in crowded fields, such as those close to the centers of rich LMC star clusters. In these regions, SVA1 samples less than 50\% of the objects detected by \textsc{daophot}, even at bright magnitudes (g $\textless$ 19). Inside rich cluster cores, the SVA1 incompleteness is large enough to cause local holes in the surface density distribution of point sources on the sky. This result is robust to the way we separate stars from galaxies.

\end{enumerate}

With these results in mind, we decided that we could not use the SVA1 catalogs inside and around the LMC clusters. However, we used SVA1 coadd images to identify cluster candidates and its catalogs to fit zeropoints to \textsc{daophot} photometry. We also decided to use three bands (two colors) instead of two, to better describe the stellar locus of each cluster candidate. For reference, a comparison of both software packages (\textsc{sextractor} with \textsc{psfex} and \textsc{daophot} with \textsc{allstar}) can be found in~\citet{2013PASP..125...68A}. In \S \ref{daophotpipe} we describe in detail the data reduction and analysis based on our \textsc{daophot} pipeline.

\subsection{\textsc{DAOPHOT} data reduction}
\label{daophotpipe}

As mentioned earlier, the SVA1 catalog is incomplete in crowded regions, such as inside LMC star clusters. To bypass this problem, we used the SVA1 coadd image products in the vicinity of each of our candidate clusters as inputs for our own photometric extraction based on \textsc{daophot}. For each object we made an image cutout with 6.75 arcmin on a side, visually centered on the candidate, in three bands: $g$, $r$ and $i$.

\begin{figure*}
\centering
\includegraphics[width=130mm,clip=]{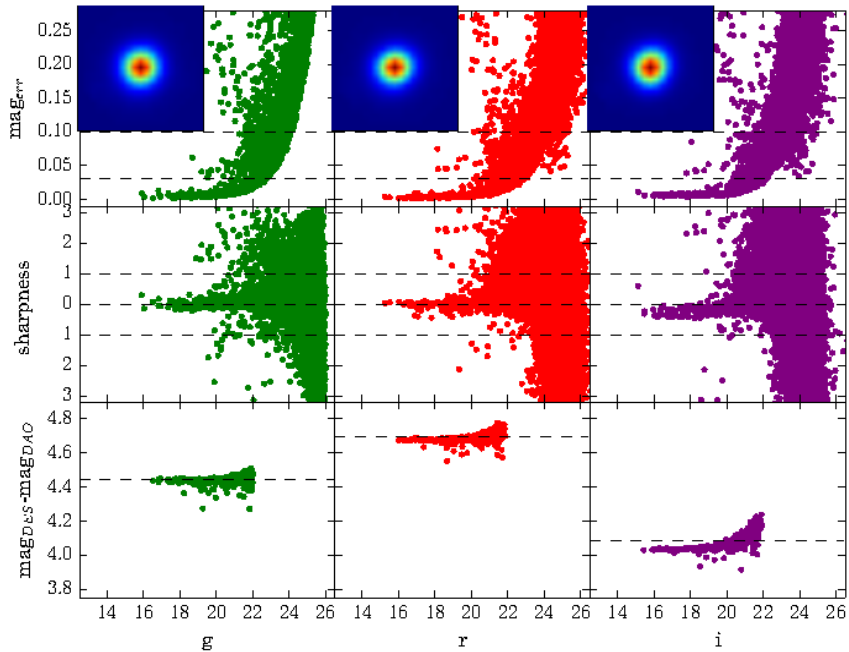}
\caption{Diagnostic plot for the image reduction around our LMC cluster candidate 67. The three columns refer to $g$, $r$ and $i$ filters, from left to right. The upper row shows the \textsc{daophot} magnitude errors increasing towards fainter stars. The upper dashed line is the maximum magnitude error adopted to classify an object as a star. The lower dashed line is the cut-off used for calibration purposes. The model PSF images are shown at the top left of these panels. In the middle row, the sharpness parameter is plotted against the magnitudes. Again, the dashed lines limit the assumed stellar locus. The lower row shows the magnitude differences ($mag_{DES}-mag_{DAO}$) for classified stars that were used for calibration (see text).}
\label{diagnostic}
\end{figure*}

The \textsc{iraf/daophot} routines were combined into a pipeline to reduce DES coadd cutouts (Fig.~\ref{flux}). This pipeline is a set of scripts using the tasks \emph{daofind}, \emph{phot}, \emph{pstselect}, \emph{psf}, \emph{allstar} and works as follows. First, we combined \emph{g} and \emph{r} images and ran \emph{daofind} to detect sources on these combined images; we rejected sources that were less than 3.0 $\sigma$ above the background. We then picked the 50 brightest sources to use as a starting template for PSF modelling (running \emph{psf} task). With the initial PSF model, we ran \emph{allstar} over all sources found and refined our set of PSF stars by choosing those with small enhancement above the sky background ($\leq$ 3 counts), low magnitude error ($\leq$ 0.01 mag) and low chi-square ($\leq$ 0.2). This procedure of refining the PSF after an initial \emph{allstar} run is meant to make the process subject to minimal human interference (although some human decision were still needed at times). The full processing requires about 20 minutes for each run on one image using a single-core desktop computer. \textsc{python} and \textsc{fortran} routines were made to select the stars and convert their $\alpha$ and $\delta$ coordinates to/from x and y coordinates. Using a program similar to the \emph{join} Linux command, we then composed the \textsc{daophot} catalogs from each filter. We then determined \emph{g}, \emph{r} and \emph{i} zeropoints, by taking DES catalog photometry as reference, and created the final \textsc{daophot} catalog. The stars used for this calibration were those that satisfied the following criteria: $magerr_{DAO} \leq 0.03$ and $\mid{sharp}\mid \leq 1.00$ in $g$, $r$ and $i$ filters. The corresponding sources in SVA1 had to have $\mid spread\_model \mid \leq 0.002$ and magnitude error $\leq 0.03$. The standard deviation for these difference ($\mid mag_{DAO}-mag_{SVA1} \mid$) has typical values $\leq$ 0.02.

After the reduction, diagnostic plots were made to evaluate the \textsc{daophot} catalog completeness level and to make reduction sanity checks. An example of these diagnostic plots is shown in Fig. \ref{diagnostic}.

The criteria we used to define sources as stars from the \textsc{daophot} catalog were magnitude error $\leq 0.1$ and $\mid{sharpness}\mid \leq 1.00$ in \emph{g, r} and \emph{i} bands. Stars used for refining the PSF model or for calibration purposes were subject to more stringent selection criteria, as explained earlier. Our \textsc{daophot} based selection of stars is still prone to substantial contamination by galaxies. However, within a typical visual cluster radius this contamination amounts to less than 2\%.

A parallel analysis of relative completeness (SVA1 and \textsc{daophot} catalogs, using the same pipeline reduction presented here) has been reported by~\cite{2015MNRAS.449.1129B} in their sec. 3.3.

\section{Methods}
\label{methods}

\subsection{Radial density profiles}
\label{rdpfit}

We used the King azimuthally symmetric profile~\citep{1962AJ.....67..471K} as the standard model to describe the surface density profile of the star cluster candidates:

\begin{equation}
\label{2}
\rho_{cl}(r) = k\left(\frac{1}{[1+(r/r_c)^2]^{\frac{1}{2}}} - {\frac{1}{[1+(r_t/r_c)^2]^{\frac{1}{2}}}}\right)^2
\end{equation}
\\
\noindent
where $\rho_{cl}$ is the surface number density of cluster stars, $r$ is the angular distance from the center ($\alpha_c$ and $\delta_c$ in cutout), $r_c$ is the cluster \emph{core radius} and $r_t$ is the cluster \emph{tidal radius}. The parameter $k$ is the profile normalization and is related to the central surface density. The King density profile is widely used for both high and low mass Galactic clusters~\citep{2013ApJ...774..151M,2012A&A...543A.156K}.

The King parameters $r_c$, $r_t$ and centroid were determined using a maximum-likelihood estimate (MLE). First, we determined the background density of stellar sources in the cutout image, $\rho_{bg}$. We do that by counting stars farther than two times a visually determined cluster radius and dividing this number count by the corresponding area. After we determined the background density, we estimated the number of stars $(N_{\star})$ belonging to the cluster by following the recipe by \cite{2008ApJ...684.1075M}:

\begin{equation}
\label{3N}
N_{\star} = N_{tot}-\rho_{bg} \times A
\end{equation}
\\
\noindent
where $A$ is the cutout area and $N_{tot}$ is the total number of stars in the cutout. The profile normalization constant $k$ is determined for each profile by dividing the number of cluster stars ($N_{\star}$) by the integral of the King profile from the center to $r_t$:

\begin{equation}
k = \frac{N_{\star}}{\int_0^{r_t} 2 \pi r dr (\frac{1}{[1+(r/r_c)^2]^{\frac{1}{2}}} - {\frac{1}{[1+(r_t/r_c)^2]^{\frac{1}{2}}}})^2}
\end{equation}
\\
In the fitting process, we varied the center position ($\alpha_c$ and $\delta_c$) and the parameters $r_c$ and $r_t$ (these parameters are determined from the model profile grid), evaluating the initial estimates by eye.

The likelihood that star $i$ belongs to the full model (King profile) with radii $r_c$ and $r_t$ and centered at $\alpha_c$ and $\delta_c$ and normalized to $k$ is:

\begin{equation}
\ell_i = k\left(\frac{1}{[1+(r_i/r_c)^2]^{\frac{1}{2}}} - {\frac{1}{[1+(r_t/r_c)^2]^{\frac{1}{2}}}}\right)^2 + \rho_{bg}
\end{equation}
\\
where $r_i$ is the radial distance of the given star from that model center.
\noindent
The most likely model (defined by parameters $r_c$, $r_t$ and position $\alpha_c,\delta_c$) is the one which maximizes the log-likelihood summed over all stars:
\begin{equation}
\log \mathcal{L}(r_c, r_t, \alpha_c, \delta_c) = \sum_{i=1}^N \log(\ell_i) 
\end{equation}
\\
where $N$ is $N_{tot}$ in equation~\ref{3N}. Notice that, in practice, stars located outide the tidal radius of each model profile contribute to the likelihood with $\ell_i = \rho_{bg}$.

\subsection{Isochrone fits}
\label{cmdfit}

We also used a maximum-likelihood approach to determine the cluster sample stellar populations (SSP): age, metallicity, distance modulus and reddening. As in the fit for the density profiles, the basic step is to measure the likelihood that each star is taken from a modelled isochrone displaced by a given distance modulus and extinction vector. In this work, we are using three magnitudes, $g$, $r$ and $i$. Thus, the distance from the isochrone in magnitude space must be evaluated in 3D space. For this purpose, we used the PARSEC isochrones~\citep{2012MNRAS.427..127B} and adopted the Galactic extinction law of~\cite{1989ApJ...345..245C}. The reddening is used to constrain extinction as measured in each magnitude. As a result of our choice, extinction is then given as

\begin{equation}
\label{9a}
A_g = 3.318 E(g-r)
\end{equation}
\begin{equation}
\label{9b}
A_r = 2.318 E(g-r)
\end{equation}
\begin{equation}
\label{9c}
A_i = 1.758 E(g-r)
\end{equation}
\\
\noindent
Notice that this choice of extinction law leads to very similar results (within the uncertainties) to a typical LMC extinction curve~\citep{2003ApJ...594..279G} for $g, r$ and $i$ DECam filters.

The isochrone fitting works as follows. Given the best-fit density profile, we first assigned a probability that star $i$ belongs to the candidate cluster as:

\begin{equation}
\label{3}
P_i^{kp} = \left(\frac{\rho_{cl}(r_i)}{\rho_{bg}+\rho_{cl}(r_i)}\right)
\end{equation}
\\
\noindent
where $\rho_{bg}$ is the background star density, as described in the previous section. We applied a threshold cut in $P^{kp} > 0.05$ to select the stars to use in the fit. We then computed the Gaussian likelihood that a given isochrone is the correct one describing this set of stars. We first determined the distance of star $i$ to isochrone $j$, $d_{ij}$. This is the minimum distance of the star in the $M$-dimensional photometric space (in this case, \emph{g}, \emph{r} and \emph{i} magnitudes) to the isochrone: 
\begin{equation}
\label{4}
d_{ij}^2 = min\left[\sum_{l=1}^M {\left(\frac{m_{li} - m_{lj}}{\sigma_{m_{li}}}\right)}^2\right]
\end{equation}
\\
\noindent
where the sum is over all photometric bands, $(m_{lj})$ is the closest isochrone magnitude in band $l$ to the observed magnitude of star $i$ ($m_{li}$), and $\sigma_{m_{li}}$ is the uncertainty in $m_{li}$. The isochrone magnitudes are already displaced by the distance modulus and reddening, both of which are free fit parameters. To avoid numerical limitations, we interpolate the isochrone points when determining $d_{ij}$, instead of using the discrete set of isochrone points.

Notice that $d_{ij}$ corresponds to the highest likelihood term that the star $i$ is drawn from isochrone model $j$, which is then given by

\begin{equation}
\label{5}
P_{ij} = \frac{1}{{(2\pi)}^{M/2}} \left(\prod_{l=1}^M \frac{1}{\sigma_{mli}} \right) \exp {\left( \frac{-d_{ij}^2}{2} \right)}
\end{equation}
\\
Finally, the logarithmic likelihood that the set of $N$ stars that satisfies the $P^{kp} > 0.05$ criterion are drawn from isochrone model $j$ can be written as

\begin{equation}
\label{6}
\log \mathcal{L}_{j} = \log \prod_{i=1}^N (P_{ij} P_i^{kp}) = \sum_{i=1}^N \log(P_{ij}) + \sum_{i=1}^N \log(P_i^{kp})
\end{equation}
\\

\subsection{Optimization methods}
\label{fitgrids}

We are dealing with a maximization problem where we want to find the model which best describes the set of likely cluster stars. Given a grid in parameter space of age, metallicity, distance and reddening, we found the peak of $\mathcal{L}$ ($\mathcal{L}_{max}$), and we probe the $2\log(\mathcal{L})$ space around this global maximum. The $2\log (\mathcal{L})$ values behave similarly to a $\chi^2$ distribution, assuming that uncertainties have Gaussian (or similar) behaviour close to the peak. In this way we estimate the $k\sigma$ confidence level in the same manner as the $\chi^2$ distribution~\citep[sec. 10.3]{Lupton1993} but considering the covariance among parameters using the profile likelihood technique~\citep{S2000,F1956}. For example, when determining the uncertainty in the age we scan the likelihood in age, maximizing with respect to the other parameters. Thus, the $k\sigma$ confidence intervals are determined when the log of the profile likelihood drops by $(k^2)/2$ from its maximum value. The uncertainties quoted in this work correspond to 1$\sigma$ confidence level (68\%).

We test our profile and isochrone fitting in simulated star clusters in Sec.~\ref{clussimul}. The next section presents the model grids to do that.

\subsection{Model grids}
\label{modelgrids}

The likelihood optimization methods outlined in the previous sections require a grid of models, both for the structure (profile fitting) and for the other physical parameters (CMD fitting).

The LMC clusters vary in size by about an order of magnitude, as will be shown later in this paper. Therefore, it was not possible to use a single grid in $r_c$ and $r_t$ for all clusters. Furthermore, initial guesses from visual estimates were not always useful in constraining an optimal range in core and tidal radii. As a result, the profile fits were carried out interactively, with a size grid that varied from one cluster to the other. The central positions were allowed to vary within $\pm r_t / 4$ as taken from the previous iteration, until convergence. We define convergence to occur when the algorithm indicates a likelihood maximum close to the center of the parameter's range. For some candidates, the range in $r_t$ continuously increased and reached the limit of the cutout (much greater than the visual radius), which means the set of stars does not present a clear overdensity. These cases were removed from our catalog.

The CMD fits were carried out with a fixed initial set of isochrones from which we built a likelihood map for each cluster. This initial grid covers the range $8.12 \leq \log[\rm age(yrs)] \leq 10.12$ with 10 equally spaced age steps of $\Delta \log[\rm age(yrs)] = 0.2$. Metallicity varies within $0.0002 \leq \rm Z \leq 0.019$ in 23 values: $0.0002 \leq Z \leq 0.0008$, step=0.0002; $0.001 \leq Z \leq 0.019$, step=0.001. The other dimensions are $18.2 \leq (m-M)_0 \leq 18.8$ with $\Delta \rm (m-M)_0 = 0.03$ and $0 \leq \rm E(g-r) \leq 0.2$ with $\Delta E(g-r)=0.02$. This grid is the same for real and simulated clusters.

We scanned the entire initial grid, searching for a global likelihood maximum. We then defined another model grid around this maximum, which is narrower in log[age(yrs)], covering $\log[\rm age(yrs)]_{max} -0.4 \leq \log[\rm age(yrs)] \leq \log[\rm age(yrs)]_{max} +0.4$ with $\Delta \log[\rm age(yrs)] = 0.02$, where $\log[\rm age(yrs)]_{max}$ is the age corresponding to the likelihood peak in the initial grid. Metallicity in this thinner grid was restricted to $\pm 4$ steps around the maximum likelihood value in the initial grid. Reddening and distance modulus were allowed to vary by $\pm 0.05$ and $\pm 0.1$ respectively, with 10 steps in each axis, from their initial best solution.

\section{Cluster simulations}
\label{clussimul}

We first tested the fitting methods for density profiles and isochrones described above using simulated data.

\begin{figure*}
\centering
\includegraphics[width=150mm,clip=]{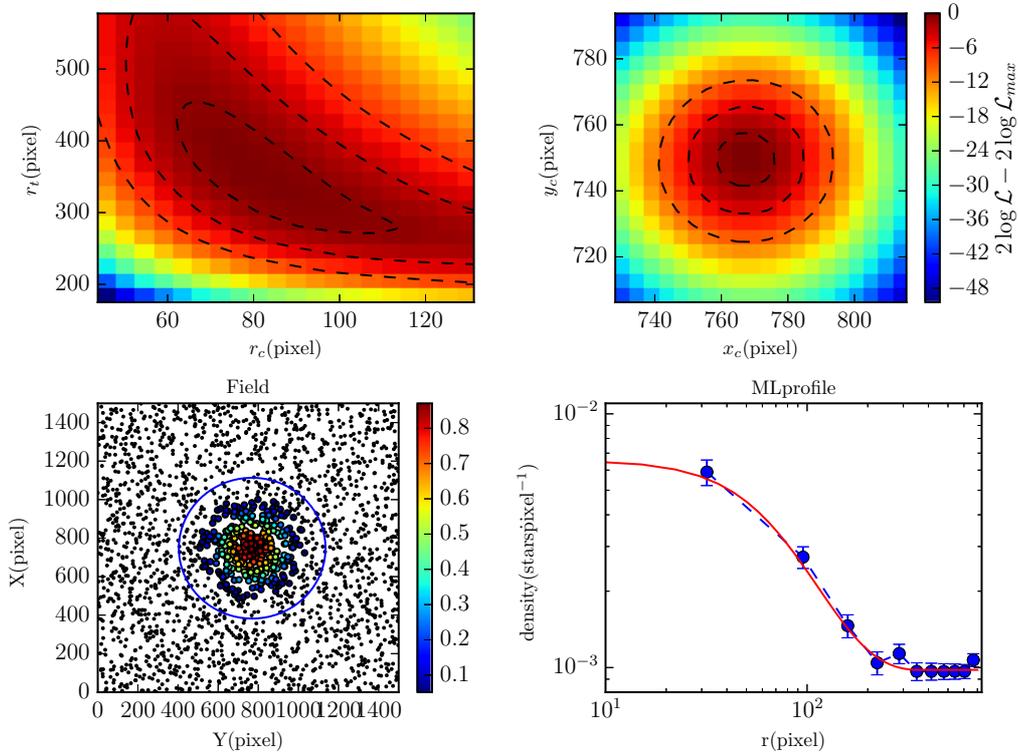}
\caption{Example of the recovery of structural parameters for a simulated cluster ($\tau$ = 1Gyr and Z = 0.010). The most likely parameters (using the DECam plate scale = 0.27 arcsec pixel$^{-1}$) are $r_c = 20.9$ arcsec, $r_t$ = 98.8 arcsec, $k = 0.14$ stars arcsec$^{-2}$. \emph{Top left:} Likelihood maps for grid in $r_c$ and $r_t$. \emph{Top right:} $2 \log \mathcal{L}$ map for center position (top left and right plots share same colorbar). Black dashed lines in both top panels show limits for 1, 2 and 3 sigma confidence levels. \emph{Bottom left:} on-sky distribution of stars, color coded according to likelihood. \emph{Bottom right:} Binned profile (blue dashed line) and best-fit model profile (solid red line).} 
\label{4profileyh}
\end{figure*}

We built artificial star clusters using \textsc{gencmd}\footnote{https:/github.com/balbinot/genCMD} and we inserted the synthetic stars into a typical LMC field. This code generates simple stellar populations ($\alpha$, $\delta$, magnitudes and magnitudes errors) given an initial number of stars, a mass function, a spatial density profile, a set of isochrones, filters and typical magnitude uncertainties. Here we used the Kroupa mass function \citep{2001MNRAS.322..231K}, King profile \citep{1962AJ.....67..471K} and PARSEC models~\citep{2012MNRAS.427..127B} for DECam filters (g, r and i). The positions of the synthetic stars are randomly chosen from the King profile for some choice of $r_c$, $r_t$ and density normalization $k$. Their magnitude are also randomly picked given the chosen isochrone model (age, metallicity, distance and reddening) and mass function. We used the same extinction law as cited above for DES bands and a typical magnitude error curve for each filter.

We simulated star clusters at a fixed distance modulus of $\rm (m-M)_0 = 18.5$ and no reddening. The simulated clusters were all cut at $g \leq 24$ and $\sigma_{g,r,i} \leq 0.1$. These photometric uncertainties were assigned to the synthetic stars following empirical curves as a function of \emph{gri} magnitudes taken from \cite{2015MNRAS.449.1129B}. For simplicity, we did not simulate unresolved binaries. We also used a fixed concentration parameter typical of LMC clusters for all the simulations, $\log_{10}(r_t/r_c) = 0.6$~\citep{2011AJ....142...48W}. 

We produced sets of clusters with different richness levels, varying the number of stars but keeping tidal radius constant (for a same set of simulations). Each set had six clusters, resulting from the combination of two bins in metallicity (metal-poor, Z=0.0002; and metal-rich, Z=0.010) with three bins in age (young, 1Gyr; intermediate age, 5Gyr; and old, 10Gyr). 

\begin{figure}
\centering
\includegraphics[width=130mm,clip=]{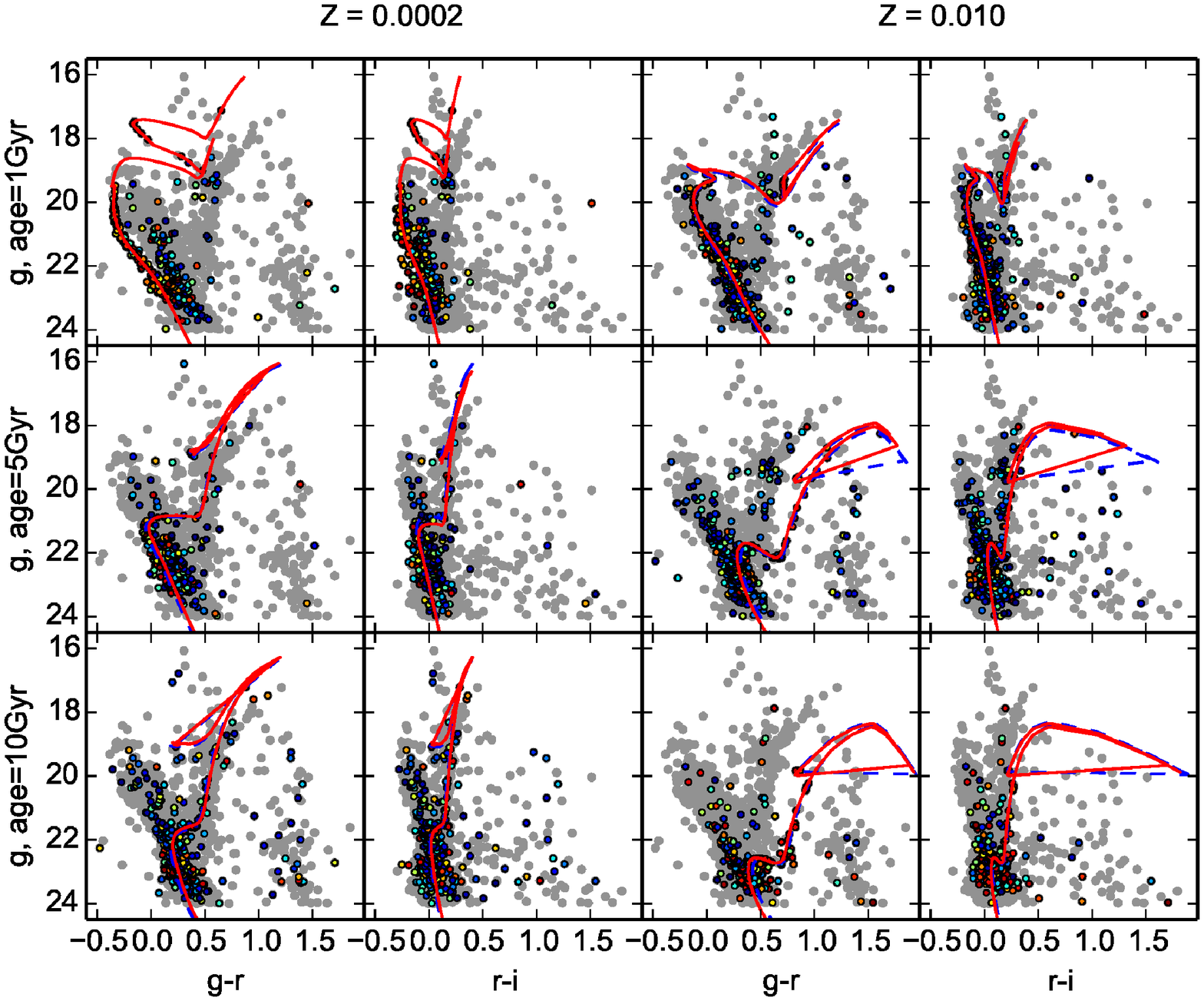}
\includegraphics[width=130mm,clip=]{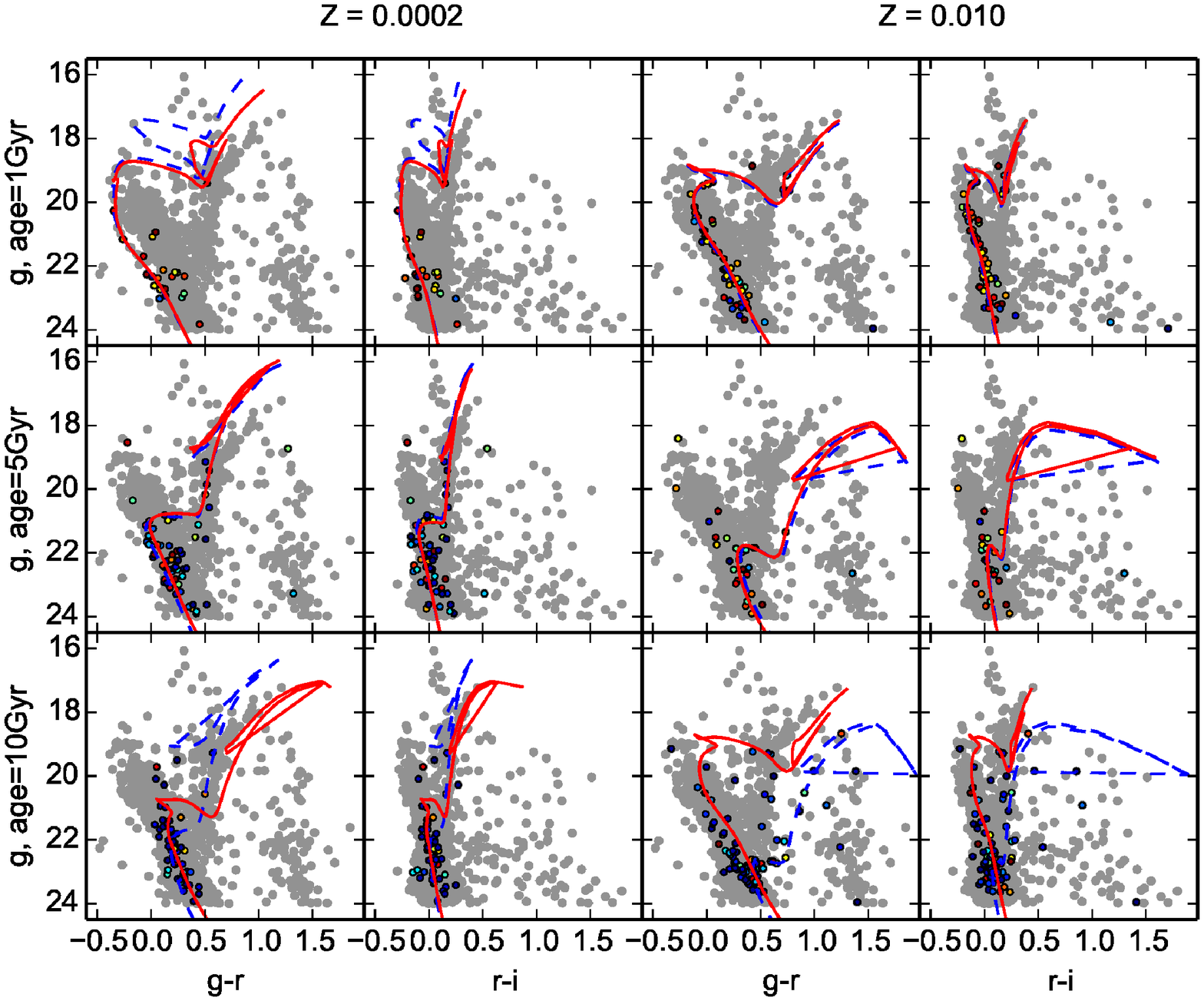}
\caption{Synthetic CMDs for six clusters inserted in a LMC field in three bands ($g$, $r$ and $i$). The clusters are combinations of young (1Gyr), intermediate age (5Gyr) and old ages (10 Gyr) with two metallicities (Z=0.0002 and Z=0.010). Stars whose likelihood to belong to a cluster is greater than 0.05 are shown in colour (and black outline), while background stars are shown in grey. The generating isochrone (dashed blue line) and the recovered isochrone (solid red line) are overlaid to the data. \emph{Top:} Clusters with 88-151 stars. \emph{Bottom:} clusters with 21-38 stars. The method fails to recover cluster parameters of old and sparse clusters (clusters with 22 and 36 stars respectively). Young and intermediate age clusters (which represent the majority of LMC clusters sample) are well recovered even for a small number of stars.}
\label{simcmd}
\end{figure}

All clusters in a set were simulated with a variable number of input stars in \textsc{gencmd}. The final number depends on several factors, including distance, magnitude and colour cut-offs, photometric errors, age and metallicity. A first set was run for clusters with a total of 1200-5400 input stars following a King profile with $r_t = 94$ arcsec. The number of stars in the output varied from 88 to 151. Fig.~\ref{4profileyh} presents the results for application of the method for a simulated cluster with $\tau =$ 5 Gyr, Z = 0.010, $r_c =$ 20 arcsec, $r_t =$ 94 arcsec and 137 stars, centered in the image center. The recovered structural parameters are $r_c$ = 20 arcsec and $r_t$ = 96 arcsec and the best fit center is offsets $\Delta x$=4.3 arcsec and $\Delta y$=0.6 arcsec, using the method described above (Sec.~\ref{methods}). The method is also very efficient in recovering the input isochrone (Fig.~\ref{simcmd}, upper panels).

To probe the limitations of the methods to cluster richness we simulated additional, poorer clusters. One of these runs simulated six small clusters ($r_t = 48 \rm arcsec$), with 21-38 stars in the output, following the same ratio $\rho_c/N$ ($\rho_c$ = central density and $N$ the cluster total star count). As shown in the bottom panel of Fig.~\ref{simcmd}, the method fails to correctly recover the old clusters (which have respectively 22 and 36 stars for the low and high metallicities). 

\begin{figure}
\centering
\includegraphics[width=130mm,clip=]{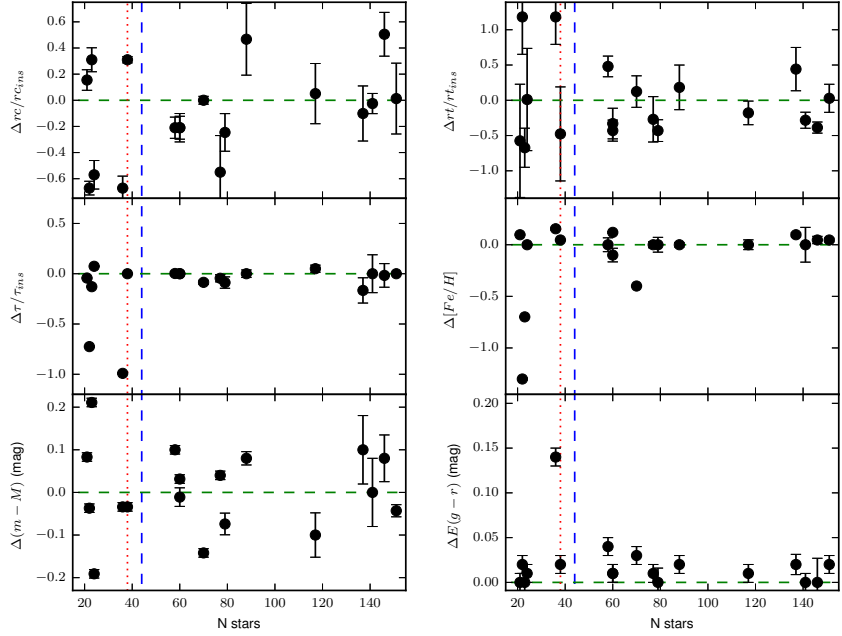}
\caption{Errors in parameter recovery as a function of number of stars in simulated clusters. \emph{Top panels:} relative errors in core radius and tidal radius; \emph{middle panels}: relative errors in age and absolute error in metallicity; \emph{bottom panels}: errors in distance modulus and E(g-r), in magnitude units. The dashed blue line shows the adopted minimum cluster richness of 44 stars in the actual LMC sample.}
\label{simpar}
\end{figure}

The errors in recovering cluster parameters for all our simulated sets are addressed in Fig.~\ref{simpar}; we show the relative parameter errors, as well as the estimated uncertainties. For the vast majority of simulated clusters, the figure shows errors of $< 20\%$ in age and metallicity, $< 10\%$ in distance, and $< 0.05$ in extinction. The structural parameters are recovered with somewhat larger errors, $\simeq 40\%$. There is a weak dependence of the error amplitude on cluster richness in most panels. This trend is more visible in the $r_t$, age, and $[Fe/H]$ plots, where errors significantly larger than those quoted above occur for simulated clusters poorer than $N \simeq 40-50$. A similar behaviour is also seen in the CMD fits presented in Fig. \ref{simcmd}, as discussed earlier. To prevent lower quality fits from contaminating our results, we adopt a somewhat arbitrary lower limit in richness of $N=44$ in our LMC sample. We note, however, that we have only 20 clusters ($\simeq$17\% of the sample) in the range 44-100 stars, and that no visible trend in age, $[Fe/H]$, distance modulus, redenning or structural parameters is seen in this richness range. This renders the results of our upcoming analysis (\S \ref{clusanalysis}) robust to the exact richness threshold adopted.

\section{DES-SV data}
\label{truedata}

In this section we show the results of our profile and isochrone fitting methods to the sample of LMC clusters. We first carried out the profile fit as described in \S \ref{rdpfit}. Only clusters with a $\mathcal{L}$ peak corresponding to $r_t < 6.5 \arcmin$ and with a minimum of $N_{star} =$ 44 member stars were used in the subsequent analyses. The upper bound in the tidal radius is the size of the coadd image cutout around each cluster candidate. The only exception to this rule is the Reticulum Cluster, whose tidal radius is larger than the image cutout size but was kept in our sample for comparison with results from the literature (\S \ref{literanalysis}). The richness criterion is guided by our simulation results, as discussed in \S \ref{clussimul}. Of the 255 cluster candidates mentioned in \S \ref{lmcclus}, only 121 had their structural parameters successfully determined by the likelihood fit. Including the Reticulum Cluster (122 candidates), the final sample available for isochrone fitting and analysis amounts to 117 clusters. Five clusters have less than the adopted minimum number of member stars and are therefore below the threshold limit.

\begin{figure}
\centering
\includegraphics[width=130mm,clip=]{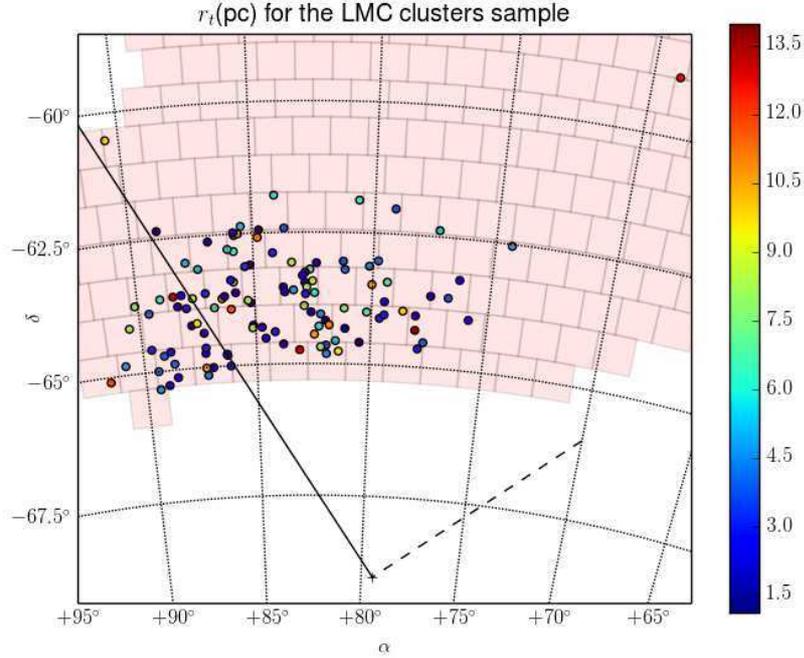}
\caption{On-sky cluster sample (circles), showing the distribution of tidal radius (in parsecs, color-coded) and the tiles sampled (complete or incompletely) in SV campaign (red boxes). The cross (bottom) indicates the LMC center. The dashed line is the LMC line of nodes and solid line is the maximum gradient distance line. The Reticulum cluster is the upper rightmost cluster.}
\label{window}
\end{figure}

Figure~\ref{window} presents their on-sky distribution colour-coded by their most likely tidal radius.

\subsection{Comparison to literature}
\label{literanalysis}

\begin{figure}
\centering
\includegraphics[width=130mm,clip=]{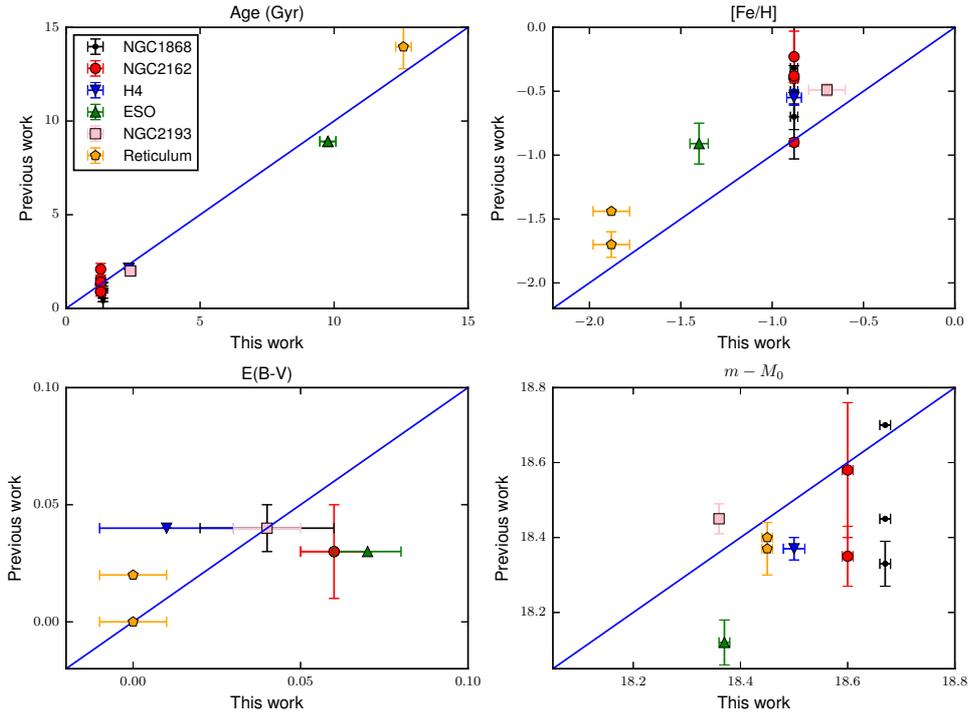}
\caption{Comparison between our fits and previous works for age (upper left), metallicity (upper right), reddening (lower left) and distance modulus (lower right). The identify line is also plotted. The uncertainties are shown in the cases they were quoted in the references.}
\label{lit}
\end{figure}

Before proceeding with a LMC cluster system analysis, we first validated the methods outlined in \S \ref{methods} with DES-SV data by comparing our LMC clusters parameters to those found in the literature. Our comparison sample is made up of six relatively rich clusters for which data of comparable or superior quality are available. For NGC 1868 and NGC 2162, there are parameter estimates from more than one source in the literature, while for Hodge 4, ESO121-03, NGC 2193 and Reticulum Cluster only one reference was found. The results are summarized in Table~\ref{literature} and Fig.~\ref{lit}.

\begin{table*}
\setlength{\tabcolsep}{2pt}
\caption{Table comparing literature data and our determinations. Numbers in parenthesis indicate the references while the last line to each cluster shows the values determined in this paper. References are: (1)~\citealt{1988AJ.....96.1383E} (CMD fitting), (2)~\citealt{1997AJ....114.1920G} (CMD fitting), (3)~\citealt{1995A&A...298...87G} (CMD fitting), (4)~\citealt{2003MNRAS.338...85M} (Surface brightness profiles), (5)~\citealt{2007AJ....134..680G} (RC and CMD fitting), (6)~\citealt{2007A&A...462..139K} (HST/CMD fitting), (7)~\citealt{2005A&A...435...77K} (HST/CMD fitting), (8)~\citealt{2006A&A...452..155K} (HST/CMD fitting), (9)~\citealt{2013AJ....145..160K} (Variable stars), (10)~\citealt{2003AJ....126.1811L} (integrated spectra), (11)~\citealt{2014ApJ...784..157L} (CMD fitting and simulations), (12)~\citealt{1991AJ....101..515O} (spectroscopy of red giants), (13)~\citealt{2013BAAA...56..275P} (CMD fitting), (14)~\citealt{2014MNRAS.444.1425P} (HST/CMD fitting).}
\label{literature}
\vspace{1em}
\centering
\begin{tabular}{cccccc}
\hline Cluster & log(age) & [Fe/H] & E(g-r) & $(m-M)_0$ & $r_c(arcsec)$ \\
\hline
NGC 1868 & $8.87\pm0.10$ (3) & & & & $6.67$ (4) \\
         & $8.74\pm0.30$ (1) & $-0.50\pm0.20$ (12) & & & \\
         & $8.95\pm0.03$ (7) & $-0.40\pm0.10$ (7) & & & \\
         & $8.97\pm0.04$ (10) & $-0.32\pm0.71$ (10) & & & \\
         & $8.95$ (8) & $-0.38$ (8) & $0.00$ (8) & 18.70 (8) & \\
         & $8.93$ (11) & $-0.38$ (11) & $0.04$ (11) & 18.45 (11) & \\
         & $9.05\pm0.03$ (6) & $-0.70\pm0.10$ (6) & $0.04\pm0.01$ (6) & $18.33\pm0.06$ (6) & \\
         & $9.14\pm0.01$ & $-0.88\pm0.02$ & $0.04\pm0.02$ & $18.67\pm0.01$ & 99 \\
\hline
NGC 2162 & $8.95\pm0.10$ (3) & & & & $10.13$ (4) \\
         & $9.20\pm0.12$ (14) & $-0.40$ (13) & & & \\
         & $9.32\pm0.06$ (10) & $-0.90\pm0.03$ (10) & & & \\
         & $9.11^{+0.12}_{-0.16}$ (2) & $-0.23\pm0.20$ (13) & & & \\
         & $9.10\pm0.03$ (6) & $-0.38$ (6) & $0.03\pm0.02$ (6) & $18.35\pm0.08$ (6) & \\
         & $9.15$ (5) & $-0.46\pm0.07$ (5) & $0.03$ (5) & $18.58\pm0.18$ (5) & \\
         & $9.11\pm0.01$ & $-0.88\pm0.01$ & $0.06\pm0.01$ & $18.60\pm0.01$ & 55 \\
\hline
Hodge 4 & $9.33$ (5) & $-0.55\pm0.06$ (5) & $0.04$ (5) & $18.37\pm0.03$ (5) & $15.2$ (4) \\
        & $9.37\pm0.02$ & $-0.88\pm0.04$ & $0.01\pm0.02$ & $18.50\pm0.02$ & 50 \\
\hline
ESO121-03 & $9.95$ (5) & $-0.91\pm0.16$ (5) & $0.03$ (5) & $18.12\pm0.06$ (5) & \\
        & $9.99\pm0.01$ & $-1.40\pm0.05$ & $0.07\pm0.01$ & $18.37\pm0.01$ & \\
\hline
NGC 2193 & $9.30$ (5) & $-0.49\pm0.05$ (5) & $0.04$ (5) & $18.45\pm0.04$ (5) & \\
        & $9.38\pm0.01$ & $-0.70\pm0.04$ & $0.03\pm0.01$ & $18.36\pm0.01$ & \\
\hline
Reticulum Cluster & & $-1.70\pm0.10$ (11) & $0.00$ (11) & $18.37\pm0.07$ (11) & \\
        & $10.11-10.18$ (9) & $-1.44$ (9) & $0.016$ (9) & $18.40$ (9) & \\
        & $10.10\pm0.01$ & $-1.88\pm0.10$ & $0.00\pm0.01$ & $18.45\pm0.01$ & \\
\hline
\end{tabular}
\end{table*}

The results listed in Table~\ref{literature} come from a variety of different methods and data. Most of them are based on CMD analysis, like our own. In this sense, the fits found in previous works are not necessarily more accurate than ours. In particular, none of those papers are based on the same set of PARSEC isochrones used in this work and some of them are not based on optical data, as is the case of~\cite{2007AJ....134..680G}. These issues, coupled with variations in methodology, are probably the cause of the spread in the parameters from different authors, or even among the same authors, as attested by the compilation of NGC 1868 and NGC 2162 results. Even with this spread, there is a clear correlation between our fits and literature, as one can see in Fig.~\ref{lit} for all the parameters.  Fig.~\ref{lit} testifies that there is no strong systematic trends in our determinations. Therefore, we conclude that the soundness of our methodology is further corroborated based on these real data comparisons.

For the structural parameters, we had no overlap with the large sample of clusters measured by~\cite{2011AJ....142...48W}. The only paper with a common sample is~\citealt{2003MNRAS.338...85M}, who measure surface brightness profiles compared to our star density profiles. The authors describe how it is difficult to reconcile their results with density profiles in the literature. A possible explanation for this discrepancy is incompleteness, which affects the density profile much more than surface brightness profiles, since fainter stars have low weight in surface brightness profile while in density profiles all stars have the same weight. The incompleteness severely affects crowded systems, as NGC 1868 and NGC 2162, for which we find core radii 15 and 5 times larger, respectively, than~\cite{2003MNRAS.338...85M}. For Hodge 4 (a less crowded cluster), this ratio decreases to 3 times larger. 

\begin{figure}
\centering
\includegraphics[width=130mm,clip=]{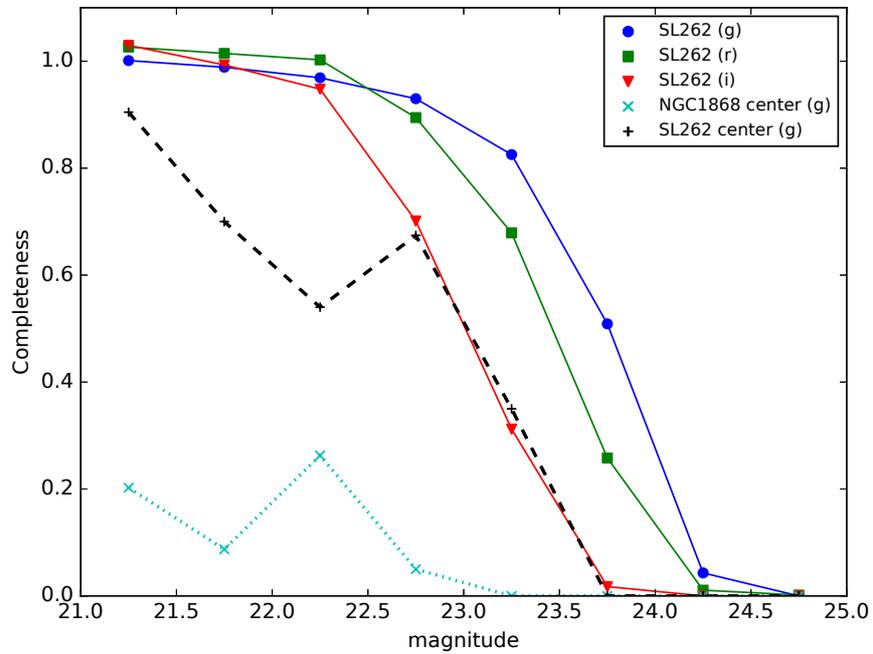}
\caption{Completeness curves integrated over all positions in the SL 262 cutout ($\simeq$ 6.6 arcmin) are shown for g, r and i bands (solid lines: circles, squares and triangles, respectively). Completeness covering central regions (a square with 53 $\times$ 53 arcsec) of SL262 (moderate crowding, dashed line and plus symbols) and NGC 1868 (severe crowding, dotted line and cross symbols) is shown just for g band as dotted lines, plus and star symbols, respectively.}
\label{comp}
\end{figure}

In Fig~\ref{comp} we show completeness curves in \emph{gri} bands integrated over the entire cutout of SL262. We also show completeness curves in the centers of moderately and severely crowded clusters, namely SL 262 and NGC 1868. For that purpose, we picked a square image (54 $\times$ 54 arcsec) centered in each cluster. For clarity curves are only presented for $g$ band, but are similar at other bands. The completeness levels shown correlate well with the strong discrepancies we find with respect to~\cite{2003MNRAS.338...85M} and suggests further that incompleteness is the dominant source of these discrepancies. Notice that rich clusters which have crowded centers make up about 15\% of our sample.

The completeness was estimated in our data by inserting $2.4 \times 10^4$ artificial stars in 10 realizations of the image in each band (g, r and i) for the two clusters cited above. We then proceeded to reduce the artificial stars in the same manner as the DES-SV data. The inserted stars vary in the range $21.0 \leq g \leq 25.0$. The criteria to recover stars are the same used to classify a source as star: $error_{g,r,i}\leq0.1$ and $\mid sharpness_{g,r,i}\mid \leq 1.0$, and a maximum deviation from initial position equal to 0.8 arcsec. 

To test whether the differences between our estimate of the core radius and those in the literature might stem from completeness, we used the same maximum likelihood method for the NGC 1868 profile, but selecting only stars with g $<$ 21.5. The corresponding core radius using the latter sample (less affected by incompleteness) decreased to 0.58 of the value initially measured (using stars with g $<$ 24). The tidal radius increased 10\% (certainly due to statistical variations, given the lower star counts in cluster outer regions). We then conclude the completeness is the main cause of difference between our determinations and those from~\cite{2003MNRAS.338...85M}.

\subsection{DES-SV LMC clusters}
\label{clusanalysis}

In this section, we analyse the distribution of LMC clusters as a function of size, age, metallicity and position in space. 

\begin{figure}
\centering
\includegraphics[width=130mm,clip=]{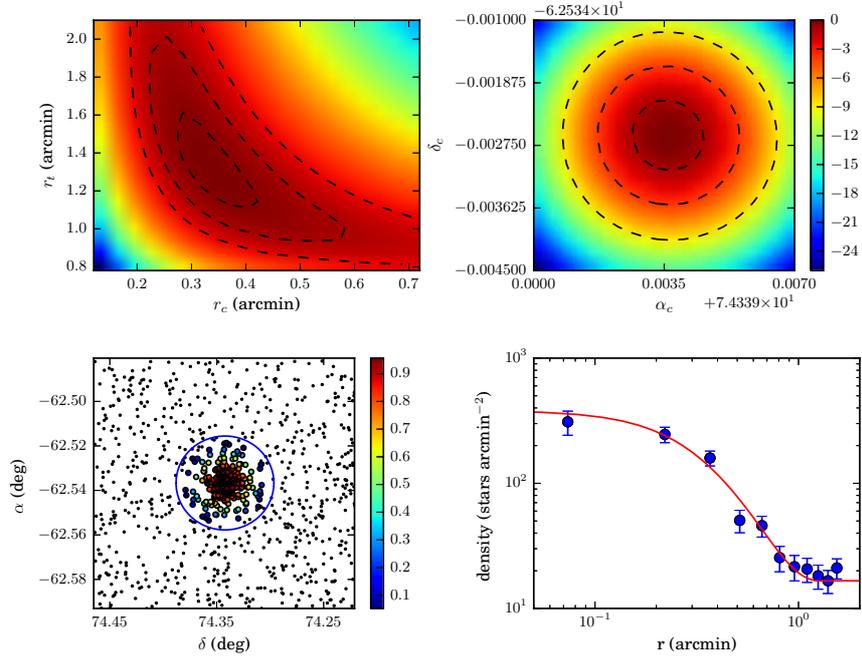}
\caption{Maximum likelihood method applied to the cluster SL126. The most likely parameters for this cluster are $r_c =$ 19 arcsec, $r_t =$ 79 arcsec, $\alpha_c = 74.3424^{\circ}$, $\delta = -62.5356^{\circ}$, $k =$ 0.14 stars arcsec$^{-2}$. \emph{Top panels:} Likelihood distribution (normalized by maximum) over $r_c$ and $r_t$ (left) and over cluster center coordinates (right). \emph{Lower left:} field stars ($P^{kp} < 0.05$, black dots) and likelihood (colorbar) that each star belongs to the cluster. \emph{Lower right:} density profile binned in 11 rings (blue dots) and the fitted density profile (solid red line).}
\label{king}
\end{figure}

\begin{figure}
\centering
\includegraphics[width=130mm,clip=]{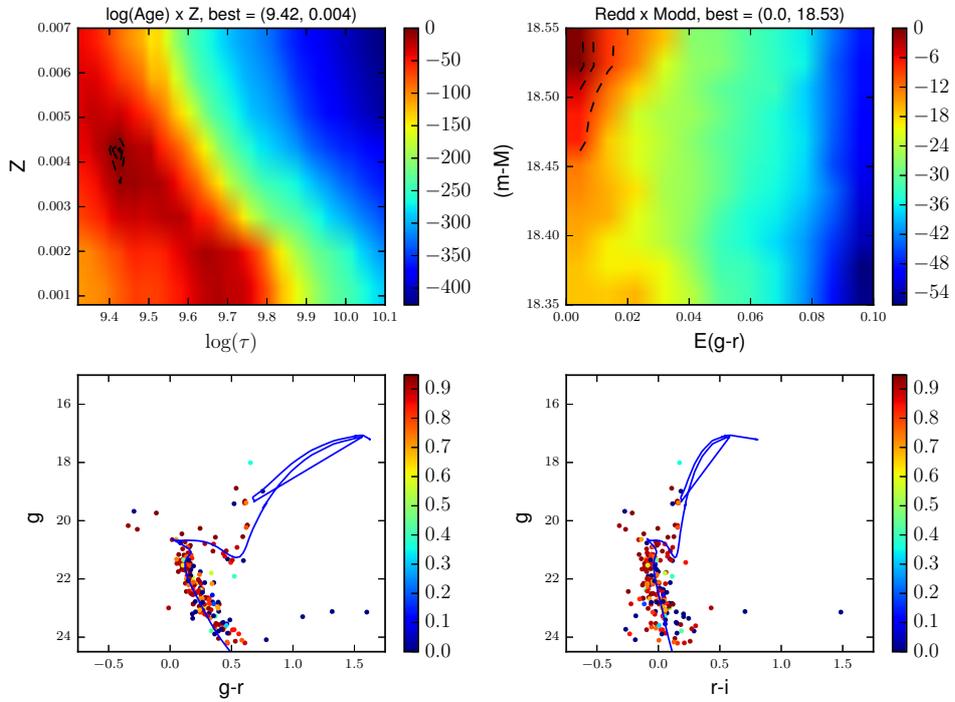}
\caption{2 log($\mathcal{L}$) distribution for age and metallicity (top left panel) and for reddening and distance modulus (top right panel) for cluster SL 126 (scaled as coded in the color bars). The bottom panels show $g~vs.~(g-r)$ and $g~vs.~(r-i)$ CMDs. Only stars with membership probability, $P^{kp} > 0.05$, are shown. The best fitting isochrone (given by MLE) is also overlaid.}
\label{DES42likelihood}
\end{figure}

Figs.~\ref{king} and ~\ref{DES42likelihood} show the results of the ML method for SL126, which is a typical cluster from our sample. Fig.~\ref{king} has a clear $\log(\mathcal{L})$ maximum for the profile and indicates that we successfully recovered the center and the structural parameters of this cluster. The CMDs in Fig.~\ref{DES42likelihood} are well described by the best-fit isochrone that resulted from the method described in \S \ref{cmdfit}.

\begin{figure}
\centering
\includegraphics[width=130mm,clip=]{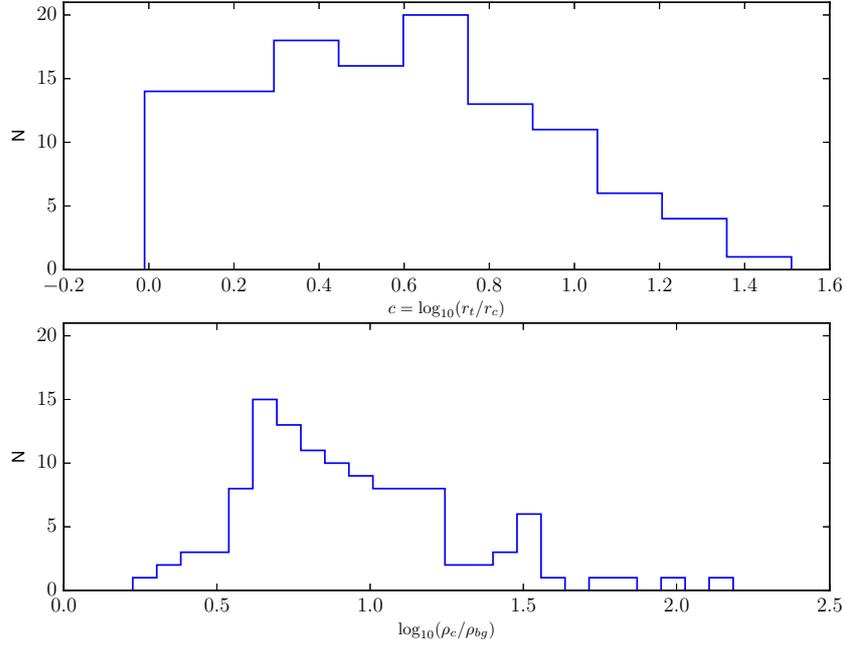}
\caption{Distribution of concentration parameter (\emph{top}) and of the logarithm of central density contrast over the background (\emph{bottom}) for the LMC clusters sample.}
\label{histconc}
\end{figure}

The concentration and $\log_{10}(\frac{\rho_c}{\rho_{bg}})$ histograms are shown in Fig.~\ref{histconc} for all clusters. The concentration distribution is very similar to the one shown by~\citealt{2011AJ....142...48W} (top panel of their Fig. 14), which is based on fits to radial luminosity profiles. Both display a broad peak around c = 0.6, and another peak at very low concentration values. In our analysis this latter peak is not as pronounced, probably attesting a strong selection bias in favour of clusters with high contrast to background (in this work), which will tend to be more concentrated at a fixed richness. Another reason for this mild discrepancy is that the authors used a radius encompassing 90\% of luminosity rather than the tidal radius. 

The central density relative background histogram has a peak $\simeq$ 3.5 times background density. A small number of clusters have densities near the background value; these are mainly located closer to the LMC center. In contrast, the high values for central densities usually occur for clusters located farther from the LMC center. The most notable cluster in this sense is the Reticulum cluster, which is located in a region where the stellar density is very low. This cluster does not have a crowded central region, allowing for good photometric parameter estimation.

\begin{figure}
\centering
\includegraphics[width=130mm,clip=]{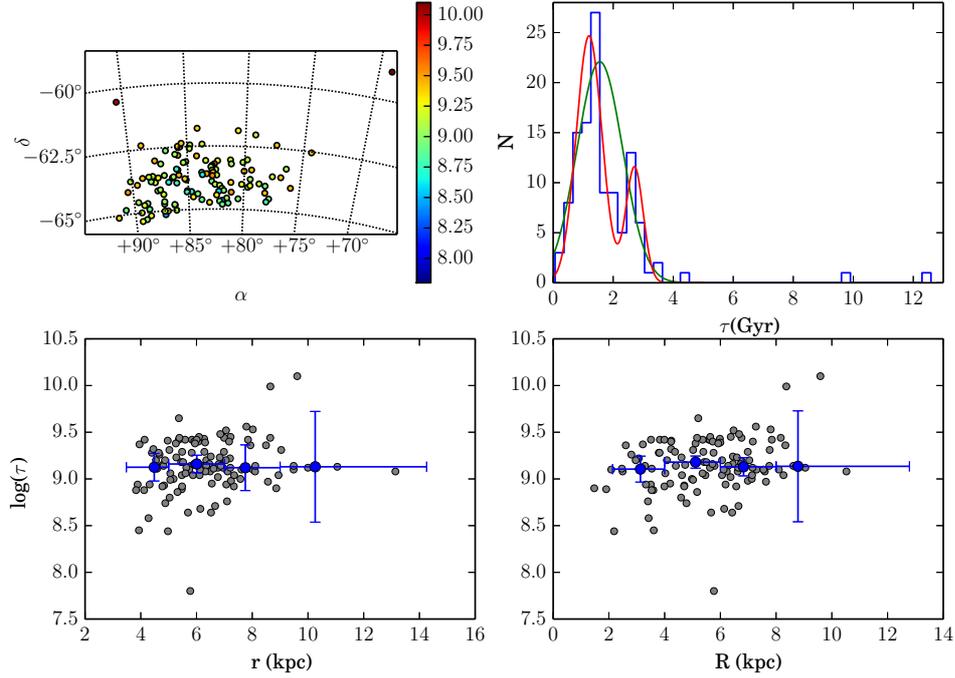}
\caption{\emph{Top left:} On-sky cluster distribution color-coded according to log(age). \emph{Top right:} Histogram of cluster ages shown along with unimodal and bimodal Gaussians fitted using the method presented in~\citet{1994AJ....108.2348A}. \emph{Bottom left:} Age versus radial distance from LMC center. Blue circles present the median age in four non-equal subsamples. Horizontal bars indicate the coordinate range. Vertical bars represent the standard deviation in age for each subsample. \emph{Bottom right:} Age versus cilindrical coordinate R.}
\label{agesample}
\end{figure}

The age distribution of the LMC clusters in our sample presents two main peaks, at $\sim$1.2 and $\sim$2.7 Gyr. This can be seen in Fig.~\ref{agesample} (top right panel). The same figure also shows ages as a function of position on the sky (top left panel) and distance to the LMC center ($r$; bottom left panel). No obvious age gradient is seen in this latter plot. 

To assess the significance of the observed bimodality, we carried out Kolmogorov-Smirnov tests based on the null-hypotheses that the observed distribution is either single or doubled peaked. We limited the test to clusters with $<$ 4 Gyr (114 clusters from the total of 117 clusters). We fitted the observed age distribution to models with one and two Gaussians. We then created 1000 random samples of ages with the same size as the real sample and following each of these models, and applied the KS test comparing each realization to the real distribution. The average result over 1000 realizations indicates that the real clusters do not originate from a unimodal distribution ($\bar p$ = 0.026 or $\simeq2 \sigma$).
Therefore, we can reject the hypothesis that the real age distribution comes from an unimodal distribution.

For the double Gaussian model, the best-fit age peaks confirm the visual estimates, corresponding to $1.2$ and $2.7$ Gyrs, respectively. And the average $p-value$ over 1000 realizations is $0.26$, showing that the observed distribution is consistent with the adopted null hypothesis in this case.

We also tried to follow the recipe in~\cite{1994AJ....108.2348A} to investigate the best-fit multi-peak model for the observed age distribution in the sample. However, our fits for one, two, and three Gaussians resulted in 
different standard deviations: 0.43 and 0.3 Gyrs for two peaks, and 0.44, 0.12 and 0.16 Gyr for three peaks. The lack of homoscedastic distributions in the current situation rendered any the results from the method by \cite{1994AJ....108.2348A} less reliable, as discussed in \cite{1994AJ....108.2348A} and \cite{1991PASP..103...95N}.

Given the observed bimodality in the age distribution, we split the sample into two classes, of clusters younger and older than 2Gyr. We calculated a mean $r$ for these two age classes, obtaining $\overline{r}=6.32$ kpc and $\overline{r} = 6.53$ kpc, respectively.

We also computed cylindrical coordinates of each cluster based on an LMC disk model from \cite{2015MNRAS.449.1129B}. These authors fit the DES-SV field stars distribution and derive the disk position angle and inclination with respect to the sky. The LMC center and heliocentric distances were kept fixed in the fit at $\alpha_0 = 79.40^\circ, \delta_0 = -69.03^\circ$ \citep{2004ApJ...601..260N} and $D_{LMC} = 49.9$ kpc \citep{2014AJ....147..122D}. To determine the cylindrical coordinates, we used the transformations presented in~\cite{2001ApJ...548..712W}. In the lower right panel of Fig.~\ref{agesample}, we plot the ages against the $R$ coordinate, along the disk plane, and again infer an essentially flat relation. The $\overline{R}$ values for our two age subsamples are 5.44 and 5.9 kpc for the younger and older clusters, respectively. 

Our sample covers an unprecedented range in distances from the LMC (from $\simeq$4kpc out to $\simeq$13kpc), reaching out to previously unexplored outer LMC regions. The clusters age distribution presented here is complementary to that shown by \cite{2013AJ....145...17P} who study field stars in regions corresponding to deprojected distances in the range from $\simeq$0.5kpc out to $\simeq$8kpc. Those authors favour an outside-in star formation in the sense that old (and metal-poor) stars tend to be located in the outer disk, whereas younger stars (also more metal-rich) tend to inhabit the inner LMC regions. This age trend with distance, however, is largely restricted to the inner 4 kpc. Beyond that, their relation between age and distance is flat, similar to what is found with the LMC clusters studied here.

\begin{figure}
\centering
\includegraphics[width=150mm,clip=]{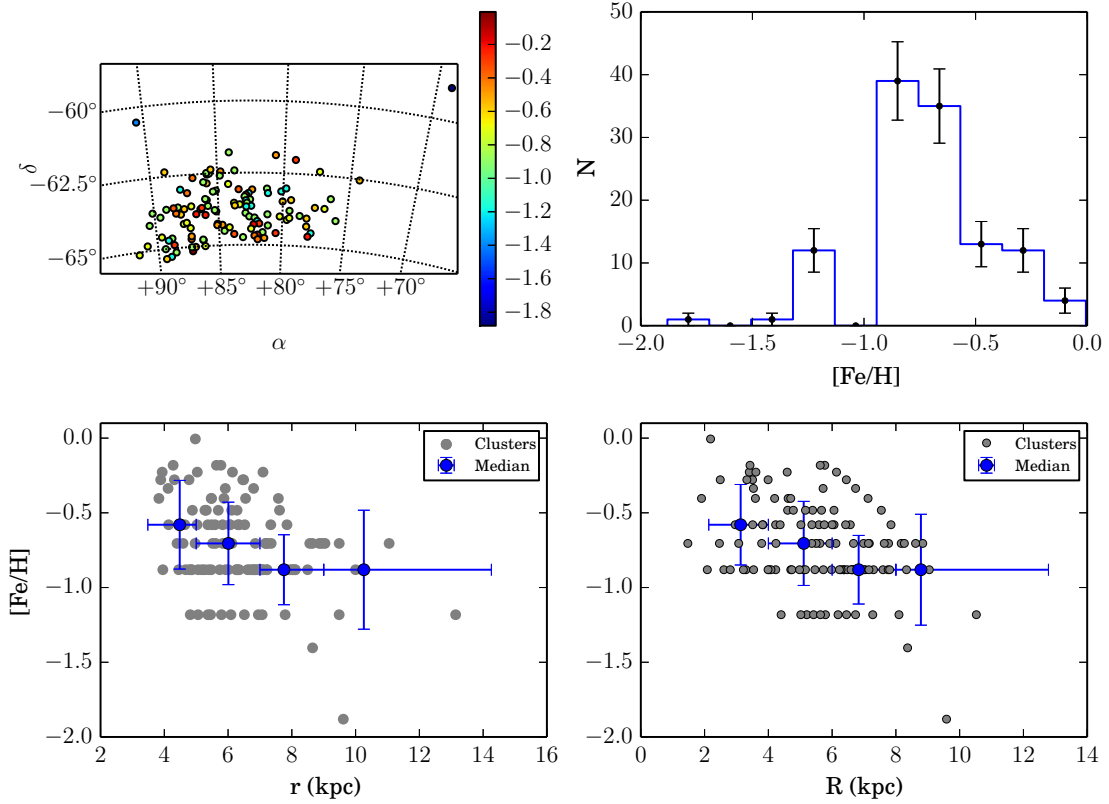}
\caption{\emph{Top left:} On sky cluster metallicity distribution (color-coded). \emph{Top right:} Metallicity histogram. \emph{Bottom left:} [Fe/H] versus radial distance from LMC center (grey circles). Filled circles present the median for metallicity, dividing the distance range in four non-equal subsamples (blue dots are placed at each subsample median distance). Vertical bars represent the metallicity disperson (standard deviation) for each subsample. Horizontal bars are the subsample range in radial coordinate (r or R). \emph{Bottom right:} Metallicity (Fe/H) for cylindrical coordinate R (radial distance projected on disk).}
\label{zsample}
\end{figure}

Fig.~\ref{zsample} shows similar plots for metallicity. Most clusters in our sample are metal poor $Z < 0.004$ ($[Fe/H] < -0.7$) as Fig.~\ref{zsample} shows in upper right panel. Unlike age, there is a clear trend in metallicity as a function of distance from the LMC center (bottom panels). The two-sided \emph{p-value} for a statistical hypothesis test (whose null hypothesis is that the slope is zero) is $10^{-5}$ for R and $5\times 10^{-5}$ for r. We conclude we can reject the null hypothesis.

The median metallicity systematically drops by a factor of $\simeq 2$ from $r = 5$kpc to $r = 10$kpc. Clusters with -1.5$<$(Fe/H)$<$-1.0 are distributed over the entire radial distance range. On the other hand, $Z > 0.005$ clusters ($[Fe/H] > -0.6$) are only found for $r < 8$kpc. A larger metallicity spread in inner fields has also been found by \cite{2013AJ....145...17P} for LMC field star population. However, for the range of distances in common with their study, our results are in disagreement in the sense that the clusters have a larger spread than the field stars at a fixed distance from the LMC center. 

\begin{figure}
\centering
\includegraphics[width=130mm,clip=]{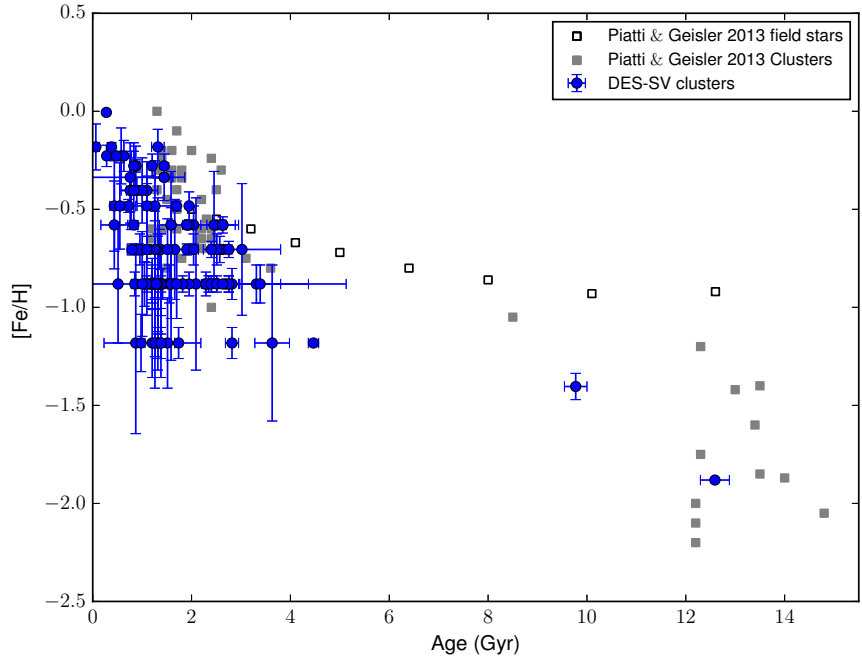}
\caption{Age-metallicity relationship for LMC clusters sample (filled circles). Open boxes (field) and filled boxes (clusters) are from \citealt{2013AJ....145...17P}~(Fig. 6 in that paper).}
\label{AMR}
\end{figure}

The age-metallicity relationship (AMR) is plotted in Fig.~\ref{AMR}, where we compare the AMR from our data (blue filled dots) to the one from \citealt{2013AJ....145...17P} (open boxes for LMC field stars and filled boxes for LMC clusters). In general the AMR from those authors corresponds to an upper envelope to the cluster AMR presented here. In particular, our sample includes young clusters with a very large range in metallicities, reaching down to $[Fe/H] \simeq -1.2$. These are mainly the clusters belonging to the 1.2 Gyr age peak. This young and metal-poor sample is consistent with a recent cluster formation epoch in the outer regions of the LMC sampled in this paper, contrasting a relative low SFR in these areas~\citep{2014MNRAS.438.1067M}. This large spread at relatively young ages contrasts with the lack of metal rich clusters at larger ages. 

Our results are in agreement with \cite{2013A&A...554A..16L}, who analyzed a sample of 15 LMC clusters spread all over the galaxy. On the other hand, the field stars AMR relation from \cite{2011AJ....142...61C} lead to higher metallicities at a fixed age than those typically inferred for our sample of outer LMC clusters, at least for ages $> 3$ Gyr. The discrepancy is not caused by the different ranges in distance to the LMC center, since most of our clusters are closer to 8 kpc, similarly to their sample. This is another example of a discrepancy between results based on field and cluster stars.

\begin{figure}
\centering
\includegraphics[width=130mm,clip=]{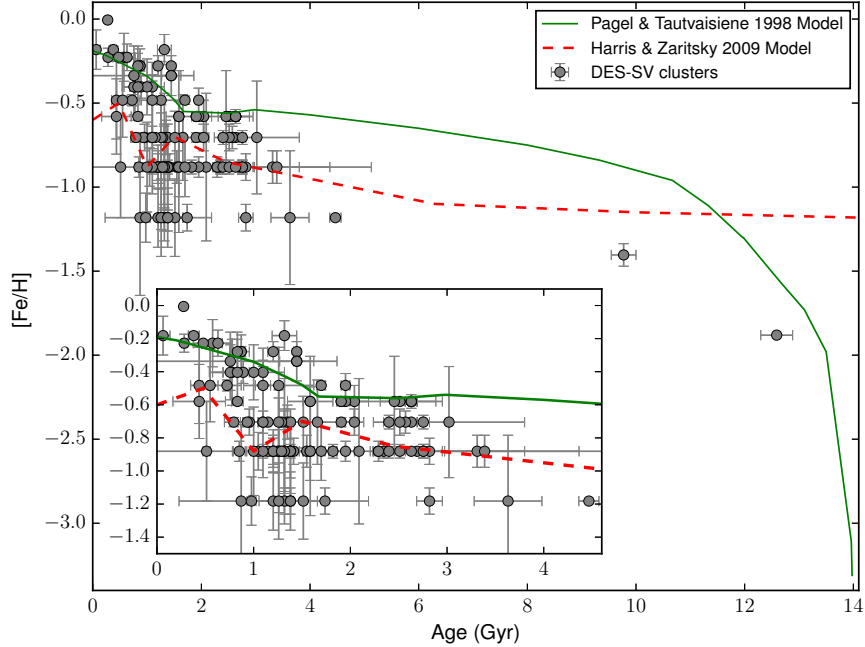}
\caption{Two models for the LMC chemical evolution history are compared to the age-metallicity relationship for outer LMC clusters (filled circles). The model from~\citet{2009AJ....138.1243H} is shown as a dashed red line, whereas the solid green line shows the~\citet{1998MNRAS.299..535P} model. \emph{Box:} Zoom in over the region indicated, showing the young and metal rich clusters compared to both models.}
\label{AMR2}
\end{figure}

In fig.~\ref{AMR2} we compare the AMR for our sample of outer LMC star clusters to the most accepted models used in literature for the LMC chemical evolution. Briefly, the model from~\cite{1998MNRAS.299..535P} is based on a bursting model assuming a constant star formation rate for clusters in the range $1.6 < \tau < 3.2$ Gyr. In that model, the metallicity increases for clusters younger than 1.6 Gyr and this feature reasonably describes the upper metallicity limit for the younger clusters studied here, as well as the old and most metal-poor clusters. The ~\cite{2009AJ....138.1243H} model is based on the StarFISH analysis code, using bright field stars.  Their results describe an initial burst of star formation and a quiescent epoch from approximately 12 to 5 Gyr ago. Star formation then resumed and has proceeded until the current time at an average rate of roughly $0.2 M_{\odot} yr^{-1}$. Among the global variations in the recent star formation rate they identify peaks at roughly 2 Gyr, 500 Myr, 100 Myr, and 12 Myr. This latter model better represents the younger clusters in our sample.~\cite{2015MNRAS.450.2122P} study a sample of clusters located in an inner LMC region and whose AMR is bracketed by the two models. Those authors argue that a combination of both models is a more adequate description of their sample than a single model. Our sample, on the other hand, shows a sizable fraction of relatively young clusters with lower metallicities than predicted by either model.

\begin{figure}
\centering
\includegraphics[width=130mm,clip=]{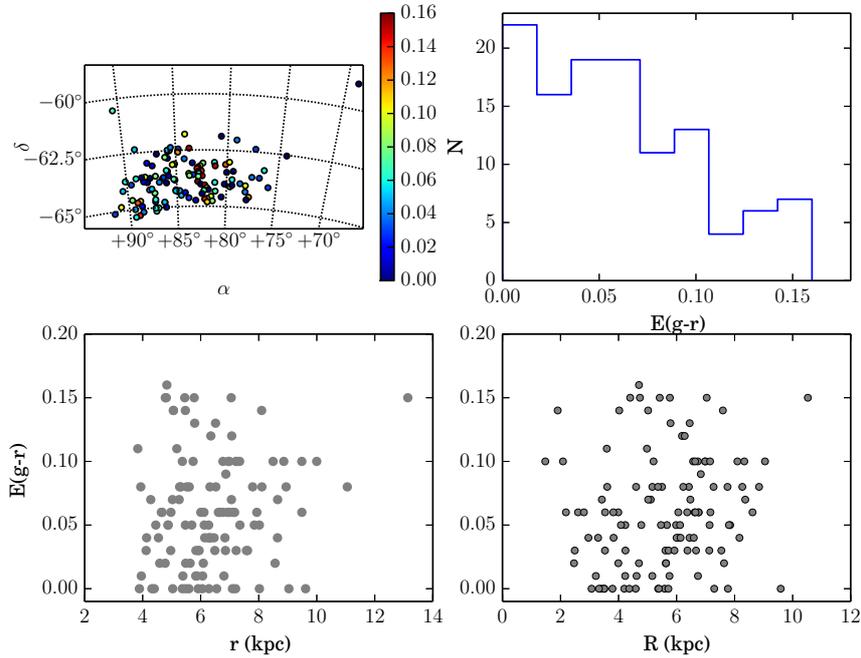}
\caption{\emph{Top left:} On sky cluster distribution color-coded according to E(g-r). \emph{Top right:} Distribution of E(g-r). \emph{Bottom:} Reddening versus radial distance from LMC center (left) and versus cilindrical coordinate R (radial distance projected on disk) at right.}
\label{reddsample}
\end{figure}

Fig.~\ref{reddsample} shows the results for the reddening values obtained from the ML fits. The sample of clusters is systematically decreasing towards larger extinction. The majority of the clusters have $E(g-r)\leq 0.08$. Using the maps from~\cite{1998ApJ...500..525S} and the reddening law from~\cite{1989ApJ...345..245C}, the typical values towards these clusters are in the range $0.04 \leq E(g-r) \leq 0.10$, which is in general agreement with the fitted values. There is no visible trend in $E(g-r)$ values with distance from the LMC center. 

\section{Discussion and Summary}
\label{summary}

Here we summarize our main results:

\begin{enumerate}[leftmargin=.75cm,labelsep=0.2cm,align=left,label=(\roman*)]
\item We scanned the DES-SV images to search for stellar overdensities in fields close to the LMC, identifying 255 cluster candidates. We cataloged and matched this sample to the star clusters catalog from~\cite{Bica2008}, adding clusters already discovered and identifying unknown clusters. We used DES-SV coadd images in \emph{g, r} and \emph{i} bands to make square cutouts (with 6.75 arcmin on a side) around each candidate.
\item We found that stellar completeness in DES-SVA1 catalog is a strong function of source density, sharply dropping to $\textless$ 0.1 for surface densities $\textgreater$ 260 stars/arcmin$^2$. The DES-SVA1 stellar sample is very incomplete in crowded fields, such as those close to centers of rich LMC star clusters, where DESDM detects less than 50\% of the objects detected by \textsc{daophot}. 
\item To reduce stellar incompleteness in crowded fields, we developed a pipeline to reduce data using \textsc{daophot}. The pipeline combines g and r images and runs PSF selection and photometry in a largely automatic way. Using stars with good photometry in both catalogs, we compared the final \textsc{daophot} magnitudes to DES-SVA1, determining a zeropoint. The agreement is very good and on average DES-SVA1 and \textsc{daophot} photometry agree within 0.02 in g and r bands, without the need for a color term. We selected stars as sources from \textsc{daophot} reductions with \emph{g,r,i} errors $< 0.1$ and $\mid sharpness_{g,r,i} \mid < 1$.
\item  We applied a maximum likelihood estimation approach to fit the stellar density profile to a King model \citep{1962AJ.....67..471K}, following a recipe similar to that of~\cite{2008ApJ...684.1075M}. This method is robust to determine center position, core and tidal radii. As output, for each cluster candidate we listed the stars within the tidal radius, along with their probability to belong to cluster candidate ($P^{kp}$). From the initial sample containing 255 cluster candidates, only 121 had their structural parameters successfully determined by the profile likelihood fit. 
\item We compared the stars with $P^{kp} > 0.05$ to \cite{2012MNRAS.427..127B} isochrones, varying reddening and distance moduli in a two-step refined grid search. The final grid allows a more refined scan of parameters space around the maximum likelihood estimate peak given by the initial grid. The uncertainties are estimated from the likelihood distribution around peak. We used the results from this method for both global and structural parameters.
\item We tested our inference methods, by inserting simulated cluster stars (using \textsc{gencmd}) in a typical LMC field and proceeding through the same steps as for real clusters. We varied the input number of simulated stars, reaching a minimal number of $\simeq 40-50$ cluster stars necessary to recover the generating isochrone. When a threshold on the minimum number of stars is applied, the LMC cluster sample decreases from 121 to 117 clusters, which is the final number of clusters used in the analysis.
\item We compared our method to results from data from the literature. The agreement is usually within the uncertainties, with a few exceptions in distance modulus, metallicity and core radii, and an excellent agreement for ages and reddenings. None of the comparison papers from the literature are based on the same set of PARSEC isochrones used in this work or even in optical data. These issues and the variations in methodology, are probably the cause of the discrepancy in the parameters from different authors. Given we are dealing with a homogeneous photometric sample, we conclude that our method is corroborated based on these real data comparisons (literature and our results).
\item The cluster age distribution presents two main peaks in 1.2 and 2.7 Gyr. We run statistical Kolmogorov-Smirnov tests to probe the bimodality significance. The tests confirm that a two-peaked age distribution is more consistent with the observed distribution, and rule out that the data follow a single peak age distribution. Splitting the sample in clusters younger and older than 2Gyr and calculating a mean radial distance to each subsample, the difference is not sensitive whether considering $\bar R$ (5.44 and 5.9 kpc) or $\bar r$ (6.32 and 6.53 kpc). This conclusion agrees to~\cite{2013AJ....145...17P}, who found a flat age-distribution for fields with radial distances $> 4$ kpc.
\item Metallicity presents a clearer radial trend, both in terms of the average metallicity or its dispersion. Metal rich clusters ([Fe/H] $> -0.6$) are concentrated within r $<$ 8kpc, whereas the most metal-poor clusters are found over the entire distance range.
\item The age-metallicity relationship for LMC clusters differs from LMC field stars AMR (from~\citealt{2013AJ....145...17P}) in the sense that the latter form an upper envelope in metallicity for age distribution. The same feature is shown in~\cite{2013A&A...554A..16L}, as the presence of young clusters (1Gyr) filling the range from most metal rich down to [Fe/H] $\simeq$ -1.2.
\item Regarding the LMC chemical evolution model, our sample shows more young metal-poor clusters than are predicted by the models of~\cite{1998MNRAS.299..535P} and~\cite{2009AJ....138.1243H}. Apart from this, our results are in broader agreement with the models.
\end{enumerate}

The discovery of 28 previously uncatalogued clusters (with profiles and models determined here) is proof that the list of LMC clusters is not complete, mainly in its less scrutinized outskirts. 

It is expected that the final DES release will provide deeper photometry, reaching further into the clusters' MS and decreasing uncertainties for their SGB and RGB stars, even the final release footprint will not cover same SV footprint. We plan to explore these data in greater depth, extracting more information about the LMC clusters, including the identification of multiple populations, modelling binarism, better constraining their shapes and assessing the correlations among different cluster properties. 

The cluster (and star) formation rate may be enhanced by tidal gravitational interactions (in this case, SMC-LMC or Galaxy-LMC). While the gravitational interaction between the Galaxy and the LMC is a controversial issue (see for example~\citealt{2013ApJ...764..161K} and~\citealt{1997NewA....2...77K}), there is a consensus for the SMC-LMC orbit period being $\simeq$2Gyr. In this sense, the age distribution of clusters as shown in Fig.~\ref{agesample} favors a strong gravitational interaction that occurred ~1.5Gyr ago, coinciding with the SMC-LMC pericentric passage predicted by the 'best orbital model' described in~\cite{2005MNRAS.356..680B}. The secondary peak at 3Gyrs is less pronounced, likely as a result of clusters disruption effects. Therefore, the relative peak heights provide an estimate of the half-life for these outer LMC clusters. As for the older clusters, they may result from processes taking place during the early evolution of the host. A problem of this scenario is that the metallicity predicted for clusters in the aforementioned simulations (where clusters formed from gas clouds pre-enriched by the formation of field stars) is higher than for LMC field stars, which is in disagreement with the results presented here. This discrepancy may be attenuated considering a inefficient gas mixing at the LMC outskirts.

Other photometric surveys are currently focused on the Magellanic Clouds, such as the Vista Magellanic Survey \citep{2015arXiv150306972C} and the Survey of the Magellanic Stellar History \citep{2015ASPC..491..325N}. Merging these data sets should result in a more complete picture of the Magellanic star clusters system and outer stellar populations. As these more remote regions should be strongly affected by the gravitational interaction involving the Galaxy, the LMC, the SMC, and other dwarf galaxies sharing similar orbits and location, these combined surveys should be useful constraints to N-body simulations describing the origin and evolution of the Magellanic System.

\section*{Acknowledgements}
\label{ack}

This paper has gone through internal review by the DES collaboration.

EdB acknowledges financial support from the European Research Council (ERC-StG-335936, CLUSTERS).

We are grateful for the extraordinary contributions of our CTIO colleagues and the DECam Construction, Commissioning and Science Verification
teams in achieving the excellent instrument and telescope conditions that have made this work possible.  The success of this project also 
relies critically on the expertise and dedication of the DES Data Management group.

Funding for the DES Projects has been provided by the U.S. Department of Energy, the U.S. National Science Foundation, the Ministry of Science and Education of Spain, 
the Science and Technology Facilities Council of the United Kingdom, the Higher Education Funding Council for England, the National Center for Supercomputing 
Applications at the University of Illinois at Urbana-Champaign, the Kavli Institute of Cosmological Physics at the University of Chicago, 
the Center for Cosmology and Astro-Particle Physics at the Ohio State University,
the Mitchell Institute for Fundamental Physics and Astronomy at Texas A\&M University, Financiadora de Estudos e Projetos, 
Funda{\c c}{\~a}o Carlos Chagas Filho de Amparo {\`a} Pesquisa do Estado do Rio de Janeiro, Conselho Nacional de Desenvolvimento Cient{\'i}fico e Tecnol{\'o}gico and 
the Minist{\'e}rio da Ci{\^e}ncia, Tecnologia e Inova{\c c}{\~a}o, the Deutsche Forschungsgemeinschaft and the Collaborating Institutions in the Dark Energy Survey. 

The Collaborating Institutions are Argonne National Laboratory, the University of California at Santa Cruz, the University of Cambridge, Centro de Investigaciones En{\'e}rgeticas, 
Medioambientales y Tecnol{\'o}gicas-Madrid, the University of Chicago, University College London, the DES-Brazil Consortium, the University of Edinburgh, 
the Eidgen{\"o}ssische Technische Hochschule (ETH) Z{\"u}rich, Fermi National Accelerator Laboratory, the University of Illinois at Urbana-Champaign, the Institut de Ci{\`e}ncies de l'Espai (IEEC/CSIC), 
the Institut de F{\'i}sica d'Altes Energies, Lawrence Berkeley National Laboratory, the Ludwig-Maximilians Universit{\"a}t M{\"u}nchen and the associated Excellence Cluster Universe, 
the University of Michigan, the National Optical Astronomy Observatory, the University of Nottingham, The Ohio State University, the University of Pennsylvania, the University of Portsmouth, 
SLAC National Accelerator Laboratory, Stanford University, the University of Sussex, and Texas A\&M University.

The DES data management system is supported by the National Science Foundation under Grant Number AST-1138766.
The DES participants from Spanish institutions are partially supported by MINECO under grants AYA2012-39559, ESP2013-48274, FPA2013-47986, and Centro de Excelencia Severo Ochoa SEV-2012-0234.
Research leading to these results has received funding from the European Research Council under the European Unions Seventh Framework Programme (FP7/2007-2013) including ERC grant agreements 240672, 291329, and 306478.

\section*{Affiliations}
{\small\it
\noindent
$^1$ Instituto de F\'{\i}sica, Universidade Federal do Rio Grande do Sul, 91501-900 Porto Alegre, RS, Brazil \\ $^2$ Laborat{\' o}rio Interinstitucional de e-Astronomia - LIneA, Rua Gal. Jos{\' e} Cristino 77, 20921-400, \\ Rio de Janeiro, RJ, Brazil \\ $^3$ Department of Physics, University of Surrey, Guildford GU2 7XH, UK \\ $^4$ Observat{\' o}rio Nacional, Rua Gal. Jos{\' e} Cristino 77, Rio de Janeiro, RJ - 20921-400, Brazil \\ $^5$ Fermi National Accelerator Laboratory, P. O. Box 500, Batavia, IL 60510, USA \\ $^{6}$ Kavli Institute for Particle Astrophysics \& Cosmology, P. O. Box 2450, Stanford University, \\ Stanford, CA 94305, USA \\ $^{7}$ SLAC National Accelerator Laboratory, Menlo Park, CA 94025, USA \\ $^8$ Cerro Tololo Inter-American Observatory, National Optical Astronomy Observatory, \\ Casilla 603, La Serena, Chile \\ $^9$ Department of Physics \& Astronomy, University College London, Gower Street, London, WC1E 6BT, UK \\ $^{10}$ CNRS, UMR 7095, Institut d'Astrophysique de Paris, F-75014, Paris, France \\ $^{11}$ Sorbonne Universit\'es, UPMC Univ Paris 06, UMR 7095, Institut d'Astrophysique de Paris, \\ F-75014, Paris, France \\ $^{12}$ Department of Astronomy, University of Illinois, 1002 W. Green Street, Urbana, IL 61801, USA \\ $^{13}$ National Center for Supercomputing Applications, 1205 West Clark St., Urbana, IL 61801, USA \\ $^{14}$ Institut de Ci\`encies de l'Espai, IEEC-CSIC, Campus UAB, Carrer de Can Magrans, s/n, 08193 \\ Bellaterra, Barcelona, Spain \\ $^{15}$ Institut de F\'{\i}sica d'Altes Energies, Universitat Aut\`onoma de Barcelona, E-08193 Bellaterra, \\ Barcelona, Spain \\ $^{16}$ Excellence Cluster Universe, Boltzmannstr.\ 2, 85748 Garching, Germany \\ $^{17}$ Faculty of Physics, Ludwig-Maximilians University, Scheinerstr. 1, 81679 Munich, Germany \\ $^{18}$ Department of Physics and Astronomy, University of Pennsylvania, Philadelphia, PA 19104, USA \\ $^{19}$ Jet Propulsion Laboratory, California Institute of Technology, 4800 Oak Grove Dr., Pasadena, CA 91109, USA \\ $^{20}$ Department of Physics, University of Michigan, Ann Arbor, MI 48109, USA \\ $^{21}$ Max Planck Institute for Extraterrestrial Physics, Giessenbachstrasse, 85748 Garching, Germany \\ $^{22}$ Universit\"ats-Sternwarte, Fakult\"at f\"ur Physik, Ludwig-Maximilians Universit\"at M\"unchen, \\ Scheinerstr. 1, 81679 M\"unchen, Germany \\ $^{23}$ Department of Astronomy, University of Illinois, 1002 W. Green St., Urbana, IL 61801, USA \\ $^{24}$ National Center for Supercomputing Applications, University of Illinois, 1205 W Clark St., \\ Urbana, IL 61801, USA \\ $^{25}$ Center for Cosmology and Astro-Particle Physics, The Ohio State University, Columbus, OH 43210, USA \\ $^{26}$ Department of Physics, The Ohio State University, Columbus, OH 43210, USA \\ $^{27}$ Australian Astronomical Observatory, North Ryde, NSW 2113, Australia \\ $^{28}$ George P. and Cynthia Woods Mitchell Institute for Fundamental Physics and Astronomy, \\ and Department of Physics and Astronomy, Texas A\&M University, College Station, TX 77843,  USA \\ $^{29}$ Department of Astronomy, The Ohio State University, Columbus, OH 43210, USA \\ $^{30}$ Department of Astronomy, University of Michigan, Ann Arbor, MI 48109, USA \\ $^{31}$ Instituci\'o Catalana de Recerca i Estudis Avan\c{c}ats, E-08010 Barcelona, Spain \\ $^{32}$ Institute of Cosmology \& Gravitation, University of Portsmouth, Portsmouth, PO1 3FX, UK \\ $^{33}$ Department of Physics and Astronomy, Pevensey Building, University of Sussex, Brighton, BN1 9QH, UK \\ $^{34}$ Centro de Investigaciones Energ\'eticas, Medioambientales y Tecnol\'ogicas (CIEMAT), Madrid, Spain \\ $^{35}$ Department of Physics, University of Illinois, 1110 W. Green St., Urbana, IL 61801, USA \\}

\appendix
\section{Clusters list}
\label{appen}

We append the list of clusters for which we fit profiles and isochrones. We match cluster centroids [column (2) and (3)] to~\cite{Bica2008} within a radius of 1 arcminute. The name of the nearest match is listed in column (1). In the case there are more than one object in the match, we assigned the nearest object. Columns (4) and (5) are the King profile core and tidal radius. Concentration [$c = \log(r_t/r_c)$] is listed in column (6). The age, metallicity, reddening and distance modulus are listed in the remaining columns, along with their associated uncertainties (for a 68\% confidence level).

\small{
\begin{landscape}
\begin{longtable}{lccccccccc}

\centering
Cluster name & $\alpha_c$ & $\delta_c$ & $r_c$(arcmin) & $r_t$(arcmin) & $c$ & Age(Gyr) & Fe/H & E(g-r) & $m-M_0$ \\ \hline
\endhead
\multicolumn{10}{c}{{Continued on next page}} \\
\endfoot
\hline
\endlastfoot
Reticulum Cluster & 69.04230 & -58.85967 & 1.94$\pm$0.01 & 3.40$\pm$0.01 & 0.24 & 12.59$\pm$0.29 & -1.88$\pm$0.10 & 0.00$\pm$0.01 & 18.45$\pm$0.01 \\
SL126,ESO85SC21,KMHK322 & 74.34242 & -62.53562 & 0.31$\pm$0.02 & 1.31$\pm$0.35 & 0.63 & 2.63$\pm$0.06 & -0.58$\pm$0.01 & 0.00$\pm$0.01 & 18.53$\pm$0.01 \\
DES001SC01 & 75.81822 & -64.02976 & 0.18$\pm$0.03 & 0.72$\pm$0.70 & 0.59 & 2.82$\pm$0.11 & -0.88$\pm$0.08 & 0.03$\pm$0.02 & 18.22$\pm$0.05 \\
SL214,LW130,ESO85SC41,KMHK512 & 76.35151 & -63.28754 & 0.50$\pm$0.05 & 0.75$\pm$0.48 & 0.17 & 1.20$\pm$0.04 & -0.88$\pm$0.09 & 0.06$\pm$0.01 & 18.25$\pm$0.01 \\
SL233,ESO85SC45,KMHK543 & 76.76607 & -63.64770 & 0.42$\pm$0.03 & 0.99$\pm$0.28 & 0.37 & 2.57$\pm$0.06 & -0.70$\pm$0.07 & 0.03$\pm$0.02 & 18.54$\pm$0.04 \\
SL262,LW146,ESO119SC40,KMHK582 & 77.34028 & -62.37969 & 0.56$\pm$0.03 & 1.49$\pm$0.23 & 0.42 & 2.63$\pm$0.06 & -0.70$\pm$0.01 & 0.03$\pm$0.01 & 18.46$\pm$0.01 \\
SL273,KMHK597 & 77.47475 & -63.63829 & 0.36$\pm$0.03 & 0.61$\pm$0.34 & 0.23 & 1.58$\pm$0.09 & -0.58$\pm$0.00 & 0.04$\pm$0.01 & 18.29$\pm$0.02 \\
DES001SC02 & 77.70962 & -64.52880 & 0.42$\pm$0.10 & 1.24$\pm$0.34 & 0.47 & 0.87$\pm$0.63 & -1.18$\pm$0.46 & 0.12$\pm$0.01 & 18.22$\pm$0.16 \\
BSDL818 & 77.92511 & -64.65323 & 0.17$\pm$0.07 & 0.74$\pm$0.80 & 0.64 & 0.58$\pm$0.21 & -0.23$\pm$0.14 & 0.00$\pm$0.10 & 18.73$\pm$0.18 \\
SL303,LW158,KMHK644 & 78.08593 & -64.30382 & 0.36$\pm$0.05 & 3.67$\pm$0.17 & 1.01 & 2.63$\pm$0.06 & -0.58$\pm$0.01 & 0.04$\pm$0.01 & 18.49$\pm$0.04 \\
DES001SC03 & 78.10203 & -64.02777 & 0.27$\pm$0.06 & 0.65$\pm$0.52 & 0.38 & 0.83$\pm$0.06 & -0.58$\pm$0.02 & 0.10$\pm$0.01 & 18.21$\pm$0.13 \\
NGC1868,SL330,LW169,ESO85SC56 & 78.65251 & -63.95714 & 1.65$\pm$0.01 & 2.35$\pm$0.01 & 0.15 & 1.38$\pm$0.04 & -0.88$\pm$0.02 & 0.04$\pm$0.02 & 18.67$\pm$0.01 \\
ESO119SC50,KMHK705 & 79.17349 & -62.02392 & 0.12$\pm$0.02 & 1.10$\pm$0.60 & 0.94 & 1.45$\pm$0.03 & -0.28$\pm$0.01 & 0.04$\pm$0.01 & 18.38$\pm$0.04 \\
SL354,LW177,ESO85SC63,KMHK712 & 79.38439 & -63.42073 & 0.89$\pm$0.08 & 1.79$\pm$0.12 & 0.30 & 2.63$\pm$0.07 & -0.58$\pm$0.04 & 0.00$\pm$0.01 & 18.38$\pm$0.01 \\
H88-257,H80F4-6 & 79.45394 & -64.04471 & 0.22$\pm$0.05 & 0.75$\pm$0.53 & 0.52 & 1.66$\pm$0.37 & -0.70$\pm$0.25 & 0.00$\pm$0.03 & 18.72$\pm$0.01 \\
H88-258,H80F4-7,KMHK720 & 79.46527 & -63.79510 & 0.39$\pm$0.05 & 0.71$\pm$0.48 & 0.26 & 1.05$\pm$0.07 & -0.88$\pm$0.11 & 0.06$\pm$0.01 & 18.25$\pm$0.07 \\
SL372,LW180,KMHK730 & 79.65956 & -64.11218 & 0.71$\pm$0.13 & 1.06$\pm$0.46 & 0.17 & 1.35$\pm$0.08 & -0.70$\pm$0.11 & 0.08$\pm$0.06 & 17.99$\pm$0.02 \\
NGC1900,SL376,LW184,ESO85SC68, & 79.74075 & -63.02412 & 0.49$\pm$0.05 & 1.30$\pm$0.04 & 0.43 & 1.32$\pm$0.09 & -1.18$\pm$0.09 & 0.10$\pm$0.01 & 18.27$\pm$0.01 \\
SL388,LW186,ESO85SC72,KMHK773 & 80.02339 & -63.48033 & 0.54$\pm$0.03 & 2.55$\pm$0.14 & 0.68 & 2.63$\pm$0.06 & -0.58$\pm$0.01 & 0.00$\pm$0.01 & 18.60$\pm$0.02 \\
OHSC10,KMHK782 & 80.16200 & -63.13500 & 0.17$\pm$0.04 & 1.30$\pm$0.56 & 0.89 & 1.38$\pm$0.03 & -1.18$\pm$0.17 & 0.13$\pm$0.01 & 18.37$\pm$0.01 \\
SL401,LW190,KMHK791 & 80.20939 & -64.00391 & 0.28$\pm$0.03 & 1.74$\pm$0.36 & 0.80 & 1.91$\pm$0.06 & -0.70$\pm$0.02 & 0.00$\pm$0.01 & 18.50$\pm$0.03 \\
KMHK815 & 80.49240 & -64.59899 & 0.35$\pm$0.06 & 0.36$\pm$0.14 & 0.02 & 0.87$\pm$0.08 & -0.28$\pm$0.10 & 0.00$\pm$0.02 & 18.48$\pm$0.15 \\
LW195,ESO119SC61,KMHK821 & 80.64759 & -61.87943 & 0.33$\pm$0.03 & 1.59$\pm$0.32 & 0.68 & 1.95$\pm$0.05 & -0.48$\pm$0.07 & 0.00$\pm$0.01 & 18.51$\pm$0.09 \\
DES001SC04 & 81.13215 & -64.32542 & 0.12$\pm$0.04 & 0.47$\pm$0.26 & 0.61 & 3.31$\pm$1.06 & -0.88$\pm$0.09 & 0.07$\pm$0.05 & 18.22$\pm$0.10 \\
KMHK854 & 81.17056 & -63.20352 & 0.12$\pm$0.04 & 1.46$\pm$0.60 & 1.07 & 1.20$\pm$0.15 & -1.18$\pm$0.18 & 0.15$\pm$0.02 & 17.88$\pm$0.12 \\
NGC1942,SL445,LW203,ESO85SC81, & 81.17974 & -63.93937 & 0.80$\pm$0.06 & 1.69$\pm$0.23 & 0.32 & 2.29$\pm$0.05 & -0.88$\pm$0.02 & 0.06$\pm$0.01 & 18.75$\pm$0.01 \\
SL448,LW205,ESO85SC82,KMHK859 & 81.24878 & -63.04850 & 0.28$\pm$0.04 & 1.16$\pm$0.38 & 0.62 & 1.55$\pm$0.04 & -0.88$\pm$0.10 & 0.06$\pm$0.01 & 18.27$\pm$0.01 \\
LW208,KMHK878 & 81.39492 & -64.75810 & 0.08$\pm$0.01 & 2.70$\pm$0.53 & 1.52 & 0.72$\pm$0.14 & -0.48$\pm$0.03 & 0.08$\pm$0.02 & 18.32$\pm$0.01 \\
LW212,KMHK882,BSDL1630 & 81.53731 & -64.56597 & 0.48$\pm$0.07 & 1.33$\pm$0.28 & 0.44 & 1.26$\pm$0.20 & -0.88$\pm$0.37 & 0.10$\pm$0.01 & 18.66$\pm$0.14 \\
BSDL1735 & 81.81636 & -64.26148 & 0.16$\pm$0.04 & 2.77$\pm$0.45 & 1.25 & 0.87$\pm$0.02 & -0.28$\pm$0.03 & 0.00$\pm$0.02 & 18.52$\pm$0.01 \\
DES001SC05 & 81.92198 & -64.80729 & 0.16$\pm$0.05 & 1.28$\pm$0.88 & 0.91 & 0.76$\pm$0.07 & -0.40$\pm$0.17 & 0.11$\pm$0.04 & 18.46$\pm$0.11 \\
SL484,LW216,KMHK918 & 81.94600 & -64.64997 & 0.24$\pm$0.05 & 1.04$\pm$0.54 & 0.64 & 0.28$\pm$0.03 & -0.23$\pm$0.05 & 0.08$\pm$0.01 & 18.47$\pm$0.14 \\
DES001SC08 & 81.97349 & -64.17541 & 0.39$\pm$0.02 & 0.39$\pm$0.19 & 0.01 & 1.26$\pm$0.41 & -1.18$\pm$0.24 & 0.10$\pm$0.04 & 18.08$\pm$0.12 \\
KMHK938 & 82.18239 & -64.68012 & 0.16$\pm$0.03 & 1.83$\pm$0.40 & 1.06 & 0.78$\pm$0.19 & -0.40$\pm$0.20 & 0.14$\pm$0.04 & 18.65$\pm$0.16 \\
DES001SC07 & 82.18870 & -64.05864 & 0.22$\pm$0.06 & 1.29$\pm$0.48 & 0.77 & 1.41$\pm$0.08 & -0.88$\pm$0.07 & 0.15$\pm$0.04 & 18.41$\pm$0.08 \\
HS317,KMHK947 & 82.25436 & -64.28881 & 0.12$\pm$0.04 & 1.51$\pm$0.70 & 1.12 & 0.38$\pm$0.06 & -0.18$\pm$0.02 & 0.07$\pm$0.05 & 18.51$\pm$0.01 \\
DES001SC06 & 82.37192 & -63.08330 & 0.46$\pm$0.01 & 0.48$\pm$0.32 & 0.01 & 1.70$\pm$0.04 & -0.48$\pm$0.02 & 0.08$\pm$0.01 & 18.49$\pm$0.04 \\
SL509,LW221,ESO85SC91,KMHK957 & 82.45045 & -63.64979 & 0.88$\pm$0.07 & 1.59$\pm$0.12 & 0.25 & 1.38$\pm$0.03 & -1.18$\pm$0.08 & 0.15$\pm$0.01 & 18.47$\pm$0.02 \\
SL511,LW222,KMHK959 & 82.45366 & -64.43882 & 0.26$\pm$0.07 & 3.29$\pm$0.37 & 1.10 & 2.04$\pm$0.10 & -0.70$\pm$0.08 & 0.00$\pm$0.01 & 18.10$\pm$0.01 \\
SL515,LW223,ESO85SC92,KMHK965 & 82.53406 & -63.42685 & 0.55$\pm$0.05 & 2.15$\pm$0.19 & 0.59 & 1.15$\pm$0.03 & -0.88$\pm$0.02 & 0.14$\pm$0.01 & 18.52$\pm$0.01 \\
SL525,LW225,KMHK973 & 82.59368 & -64.01776 & 0.47$\pm$0.02 & 0.49$\pm$0.40 & 0.03 & 1.26$\pm$0.49 & -0.88$\pm$0.13 & 0.00$\pm$0.03 & 18.17$\pm$0.02 \\
NGC1997,SL520,LW226,ESO86SC1, & 82.64348 & -63.20479 & 0.85$\pm$0.05 & 1.79$\pm$0.20 & 0.33 & 4.47$\pm$0.10 & -1.18$\pm$0.01 & 0.10$\pm$0.01 & 18.42$\pm$0.01 \\
SL529,LW229,ESO86SC5,KMHK992 & 82.78057 & -63.54049 & 0.41$\pm$0.05 & 2.31$\pm$0.23 & 0.76 & 2.75$\pm$0.06 & -0.88$\pm$0.04 & 0.06$\pm$0.01 & 18.25$\pm$0.02 \\
OHSC20 & 82.83336 & -63.67046 & 0.18$\pm$0.03 & 0.96$\pm$0.63 & 0.71 & 3.63$\pm$0.34 & -1.18$\pm$0.40 & 0.08$\pm$0.07 & 18.20$\pm$0.06 \\
LW230 & 82.84762 & -63.26148 & 0.30$\pm$0.04 & 0.66$\pm$0.48 & 0.35 & 0.06$\pm$0.07 & -0.18$\pm$0.12 & 0.15$\pm$0.04 & 18.37$\pm$0.16 \\
DES001SC10 & 82.85639 & -63.45356 & 0.27$\pm$0.04 & 0.50$\pm$0.44 & 0.27 & 1.51$\pm$0.15 & -1.18$\pm$0.23 & 0.15$\pm$0.06 & 18.39$\pm$0.16 \\
SL540,LW232,KMHK1003 & 82.89360 & -63.88714 & 0.16$\pm$0.04 & 2.19$\pm$0.51 & 1.13 & 0.85$\pm$0.06 & -0.88$\pm$0.05 & 0.16$\pm$0.01 & 18.43$\pm$0.19 \\
DES001SC09 & 82.95684 & -63.32252 & 0.32$\pm$0.07 & 0.84$\pm$0.56 & 0.42 & 1.20$\pm$0.18 & -0.88$\pm$0.12 & 0.14$\pm$0.03 & 18.19$\pm$0.10 \\
H4,SL556,LW237,ESO86SC9,KMHK1034 & 83.10791 & -64.73654 & 0.83$\pm$0.05 & 3.44$\pm$0.03 & 0.62 & 2.34$\pm$0.09 & -0.88$\pm$0.04 & 0.01$\pm$0.02 & 18.50$\pm$0.02 \\
DES001SC12 & 83.34019 & -63.59630 & 0.12$\pm$0.05 & 1.43$\pm$0.76 & 1.06 & 3.02$\pm$0.78 & -0.70$\pm$0.34 & 0.03$\pm$0.04 & 18.26$\pm$0.04 \\
DES001SC13 & 83.40447 & -63.07190 & 0.12$\pm$0.03 & 2.36$\pm$0.73 & 1.28 & 1.45$\pm$0.17 & -0.34$\pm$0.12 & 0.01$\pm$0.06 & 18.29$\pm$0.10 \\
DES001SC16 & 83.70146 & -62.42295 & 0.34$\pm$0.06 & 1.15$\pm$0.76 & 0.53 & 1.58$\pm$0.35 & -0.88$\pm$0.39 & 0.15$\pm$0.01 & 18.31$\pm$0.20 \\
OHSC22,KMHK1089 & 83.71419 & -63.63254 & 0.23$\pm$0.03 & 0.61$\pm$0.66 & 0.43 & 0.44$\pm$0.27 & -0.58$\pm$0.23 & 0.08$\pm$0.03 & 18.29$\pm$0.20 \\
DES001SC11 & 83.77045 & -63.53894 & 0.08$\pm$0.03 & 0.52$\pm$0.66 & 0.79 & 0.95$\pm$0.08 & -0.70$\pm$0.08 & 0.10$\pm$0.02 & 18.24$\pm$0.02 \\
DES001SC14 & 83.80342 & -64.61881 & 0.12$\pm$0.05 & 0.47$\pm$0.89 & 0.61 & 0.76$\pm$1.10 & -0.34$\pm$0.18 & 0.03$\pm$0.06 & 18.48$\pm$0.18 \\
E2,ESO120SC08,KMHK1119 & 84.09011 & -61.78861 & 0.46$\pm$0.04 & 1.74$\pm$0.32 & 0.58 & 2.29$\pm$0.05 & -0.88$\pm$0.06 & 0.10$\pm$0.01 & 18.22$\pm$0.01 \\
LW250,KMHK1129 & 84.15828 & -64.38628 & 0.37$\pm$0.04 & 0.97$\pm$0.40 & 0.42 & 0.93$\pm$0.04 & -0.70$\pm$0.03 & 0.02$\pm$0.01 & 18.13$\pm$0.01 \\
SL604,LW251,KMHK1127 & 84.19208 & -62.88430 & 0.42$\pm$0.03 & 0.99$\pm$0.30 & 0.37 & 2.75$\pm$0.08 & -0.70$\pm$0.05 & 0.04$\pm$0.01 & 18.17$\pm$0.01 \\
DES001SC15 & 84.57219 & -64.50597 & 0.67$\pm$0.17 & 0.68$\pm$0.34 & 0.00 & 2.09$\pm$3.04 & -0.88$\pm$0.44 & 0.05$\pm$0.06 & 18.44$\pm$0.18 \\
LW260,KMHK1168 & 84.72174 & -64.29217 & 0.16$\pm$0.07 & 0.65$\pm$0.69 & 0.62 & 0.44$\pm$0.09 & -0.48$\pm$0.23 & 0.05$\pm$0.04 & 18.33$\pm$0.12 \\
DES001SC17 & 84.74572 & -62.45673 & 0.32$\pm$0.01 & 0.33$\pm$0.36 & 0.01 & 1.26$\pm$0.37 & -0.48$\pm$0.23 & 0.03$\pm$0.09 & 18.26$\pm$0.02 \\
DES001SC20 & 84.79894 & -62.58597 & 0.39$\pm$0.08 & 3.07$\pm$0.55 & 0.89 & 1.38$\pm$0.03 & -0.88$\pm$0.04 & 0.12$\pm$0.01 & 18.38$\pm$0.02 \\
KMHK1195 & 85.09680 & -64.24833 & 0.61$\pm$0.02 & 0.62$\pm$0.31 & 0.00 & 1.32$\pm$0.14 & -0.18$\pm$0.09 & 0.02$\pm$0.04 & 18.33$\pm$0.10 \\
LW266,KMHK1198 & 85.13132 & -64.29848 & 0.20$\pm$0.08 & 2.12$\pm$0.70 & 1.02 & 0.55$\pm$0.09 & -0.48$\pm$0.22 & 0.05$\pm$0.03 & 18.41$\pm$0.10 \\
DES001SC19 & 85.14443 & -63.10033 & 0.15$\pm$0.03 & 0.35$\pm$0.21 & 0.37 & 1.38$\pm$0.19 & -0.88$\pm$0.26 & 0.09$\pm$0.08 & 18.30$\pm$0.01 \\
DES001SC18 & 85.17262 & -63.81872 & 0.13$\pm$0.05 & 0.53$\pm$0.49 & 0.62 & 1.32$\pm$0.30 & -0.88$\pm$0.25 & 0.10$\pm$0.01 & 18.27$\pm$0.06 \\
SL649,LW269,KMHK1214 & 85.28848 & -63.77132 & 0.31$\pm$0.04 & 2.11$\pm$0.33 & 0.83 & 2.40$\pm$0.14 & -0.88$\pm$0.04 & 0.07$\pm$0.02 & 18.41$\pm$0.02 \\
BSDL2771 & 85.36640 & -63.13242 & 0.32$\pm$0.14 & 1.12$\pm$0.75 & 0.55 & 1.32$\pm$0.18 & -0.88$\pm$0.26 & 0.10$\pm$0.08 & 18.08$\pm$0.01 \\
DES001SC22 & 85.47493 & -62.36156 & 0.13$\pm$0.03 & 1.44$\pm$0.74 & 1.03 & 2.45$\pm$0.44 & -0.58$\pm$0.27 & 0.04$\pm$0.06 & 18.42$\pm$0.18 \\
LW272,KMHK1241 & 85.60724 & -62.50170 & 0.14$\pm$0.05 & 3.08$\pm$0.63 & 1.33 & 1.95$\pm$0.09 & -0.88$\pm$0.05 & 0.10$\pm$0.01 & 18.32$\pm$0.05 \\
DES001SC21 & 85.76433 & -62.52862 & 0.20$\pm$0.06 & 1.01$\pm$0.58 & 0.70 & 2.51$\pm$0.06 & -0.70$\pm$0.08 & 0.06$\pm$0.02 & 18.30$\pm$0.09 \\
LW276,KMHK1269 & 85.80117 & -63.61729 & 0.50$\pm$0.01 & 0.51$\pm$0.22 & 0.00 & 2.75$\pm$0.06 & -0.88$\pm$0.04 & 0.01$\pm$0.01 & 18.42$\pm$0.03 \\
LW278,KMHK1265 & 85.80160 & -62.47070 & 0.26$\pm$0.03 & 0.79$\pm$0.53 & 0.49 & 2.69$\pm$0.06 & -0.88$\pm$0.02 & 0.05$\pm$0.01 & 18.42$\pm$0.01 \\
SL670,LW277,ESO86SC25,KMHK1268 & 85.80829 & -62.83391 & 0.33$\pm$0.02 & 1.59$\pm$0.29 & 0.68 & 2.51$\pm$0.06 & -0.58$\pm$0.02 & 0.01$\pm$0.01 & 18.55$\pm$0.02 \\
KMHK1278 & 85.86877 & -63.41479 & 0.13$\pm$0.04 & 2.02$\pm$0.59 & 1.20 & 1.38$\pm$0.27 & -0.88$\pm$0.16 & 0.08$\pm$0.02 & 18.20$\pm$0.12 \\
SL677,LW280,KMHK1286 & 85.98155 & -63.37342 & 0.29$\pm$0.04 & 0.86$\pm$0.44 & 0.47 & 0.51$\pm$0.04 & -0.88$\pm$0.30 & 0.06$\pm$0.02 & 18.30$\pm$0.01 \\
SL680,LW281,KMHK1290 & 86.01761 & -63.92466 & 0.31$\pm$0.12 & 3.07$\pm$0.37 & 0.99 & 0.63$\pm$0.04 & -0.23$\pm$0.08 & 0.00$\pm$0.03 & 18.45$\pm$0.11 \\
NGC2097,SL682,LW282,ESO56SC28 & 86.02948 & -62.78351 & 0.85$\pm$0.02 & 1.61$\pm$0.02 & 0.28 & 1.20$\pm$0.03 & -0.88$\pm$0.08 & 0.08$\pm$0.01 & 18.38$\pm$0.01 \\
SL689,LW284,KMHK1310 & 86.14288 & -65.00171 & 0.23$\pm$0.08 & 0.84$\pm$0.34 & 0.56 & 1.29$\pm$0.03 & -0.70$\pm$0.01 & 0.02$\pm$0.01 & 18.56$\pm$0.01 \\
SL694,LW287,KMHK1318 & 86.27804 & -63.67621 & 0.21$\pm$0.03 & 0.40$\pm$0.09 & 0.27 & 1.20$\pm$0.05 & -0.28$\pm$0.06 & 0.03$\pm$0.02 & 18.64$\pm$0.04 \\
KMHK1322 & 86.28898 & -64.79739 & 0.20$\pm$0.05 & 1.22$\pm$0.50 & 0.80 & 1.58$\pm$0.23 & -0.58$\pm$0.27 & 0.04$\pm$0.05 & 18.54$\pm$0.08 \\
SL696,LW286,KMHK1324 & 86.31240 & -64.77244 & 0.19$\pm$0.04 & 0.69$\pm$0.51 & 0.57 & 1.10$\pm$0.03 & -0.88$\pm$0.04 & 0.05$\pm$0.01 & 18.34$\pm$0.02 \\
SL701,LW289,ESO56SC30,KMHK1330 & 86.39115 & -63.71573 & 0.43$\pm$0.06 & 3.29$\pm$0.28 & 0.89 & 1.38$\pm$0.03 & -0.70$\pm$0.02 & 0.06$\pm$0.01 & 18.10$\pm$0.06 \\
BSDL2976 & 86.74397 & -63.86392 & 0.80$\pm$0.08 & 1.86$\pm$0.17 & 0.37 & 0.48$\pm$0.01 & -0.23$\pm$0.01 & 0.04$\pm$0.01 & 18.33$\pm$0.02 \\
OHSC26 & 86.84854 & -62.62215 & 0.11$\pm$0.04 & 0.54$\pm$0.30 & 0.68 & 0.89$\pm$0.07 & -0.40$\pm$0.07 & 0.00$\pm$0.07 & 18.57$\pm$0.01 \\
SL720,LW299,KMHK1373 & 86.94525 & -65.01115 & 0.44$\pm$0.02 & 0.64$\pm$0.36 & 0.16 & 2.04$\pm$0.15 & -0.58$\pm$0.10 & 0.07$\pm$0.07 & 18.37$\pm$0.01 \\
KMHK1381 & 87.03843 & -63.59845 & 0.23$\pm$0.03 & 0.93$\pm$0.06 & 0.60 & 1.15$\pm$0.03 & -0.70$\pm$0.01 & 0.08$\pm$0.01 & 18.42$\pm$0.02 \\
DES001SC23 & 87.19662 & -65.15252 & 0.33$\pm$0.07 & 1.13$\pm$0.22 & 0.54 & 0.28$\pm$0.01 & -0.01$\pm$0.01 & 0.06$\pm$0.01 & 18.59$\pm$0.01 \\
SL724,LW305,KMHK1388 & 87.22206 & -64.63312 & 0.24$\pm$0.04 & 0.94$\pm$0.35 & 0.60 & 1.10$\pm$0.08 & -0.48$\pm$0.23 & 0.06$\pm$0.04 & 18.30$\pm$0.18 \\
SL726,LW306,KMHK1390 & 87.24526 & -64.73359 & 0.34$\pm$0.04 & 0.69$\pm$0.41 & 0.31 & 1.00$\pm$0.32 & -0.40$\pm$0.17 & 0.03$\pm$0.06 & 18.20$\pm$0.13 \\
KMHK1391 & 87.25821 & -64.34838 & 0.55$\pm$0.02 & 0.57$\pm$0.35 & 0.01 & 1.51$\pm$0.20 & -0.70$\pm$0.16 & 0.08$\pm$0.02 & 18.37$\pm$0.04 \\
SL727,LW307,KMHK1393 & 87.26073 & -65.01212 & 0.35$\pm$0.05 & 2.63$\pm$0.20 & 0.88 & 1.82$\pm$0.04 & -0.88$\pm$0.05 & 0.06$\pm$0.01 & 18.55$\pm$0.01 \\
SL729,LW311,KMHK1399 & 87.33453 & -63.13470 & 0.38$\pm$0.06 & 1.42$\pm$0.48 & 0.57 & 1.10$\pm$0.03 & -0.40$\pm$0.07 & 0.01$\pm$0.01 & 18.51$\pm$0.01 \\
SL738,LW314,KMHK1417 & 87.53306 & -64.15352 & 0.21$\pm$0.05 & 2.58$\pm$0.33 & 1.09 & 1.26$\pm$0.13 & -0.70$\pm$0.15 & 0.05$\pm$0.03 & 18.33$\pm$0.01 \\
NGC2120,SL742,LW316,ESO86SC34 & 87.64599 & -63.67548 & 1.07$\pm$0.03 & 2.17$\pm$0.02 & 0.31 & 1.95$\pm$0.06 & -0.58$\pm$0.01 & 0.00$\pm$0.01 & 18.50$\pm$0.01 \\
LW319,KMHK1439 & 87.77290 & -64.18743 & 0.44$\pm$0.02 & 0.47$\pm$0.14 & 0.03 & 1.10$\pm$0.05 & -0.70$\pm$0.07 & 0.05$\pm$0.01 & 18.24$\pm$0.03 \\
LW320,KMHK1447 & 87.84075 & -62.99215 & 0.31$\pm$0.03 & 1.39$\pm$0.30 & 0.66 & 2.82$\pm$0.15 & -1.18$\pm$0.09 & 0.02$\pm$0.01 & 18.43$\pm$0.01 \\
LW323,KMHK1455 & 87.93822 & -63.86562 & 0.59$\pm$0.07 & 0.87$\pm$0.40 & 0.17 & 1.91$\pm$0.04 & -0.58$\pm$0.01 & 0.01$\pm$0.01 & 18.38$\pm$0.01 \\
OHSC27,KMHK1469 & 88.12257 & -63.60065 & 0.70$\pm$0.06 & 0.79$\pm$0.38 & 0.05 & 1.32$\pm$0.21 & -0.88$\pm$0.44 & 0.10$\pm$0.08 & 18.27$\pm$0.20 \\
KMHK1484 & 88.29843 & -63.80691 & 0.48$\pm$0.03 & 0.49$\pm$0.30 & 0.00 & 0.83$\pm$0.30 & -0.40$\pm$0.20 & 0.02$\pm$0.10 & 18.47$\pm$0.16 \\
SL768,LW326,ESO86SC38,KMHK1493 & 88.47378 & -63.61544 & 0.68$\pm$0.05 & 3.62$\pm$0.05 & 0.72 & 1.45$\pm$0.03 & -0.34$\pm$0.01 & 0.00$\pm$0.01 & 18.40$\pm$0.01 \\
DES001SC26 & 88.55920 & -65.14752 & 0.31$\pm$0.09 & 0.79$\pm$0.49 & 0.41 & 1.70$\pm$0.13 & -0.88$\pm$0.17 & 0.04$\pm$0.05 & 18.25$\pm$0.10 \\
DES001SC25 & 88.65418 & -64.88716 & 0.11$\pm$0.04 & 1.12$\pm$0.75 & 1.02 & 0.83$\pm$0.09 & -0.28$\pm$0.12 & 0.02$\pm$0.04 & 18.48$\pm$0.14 \\
LW332,KMHK1518 & 88.77851 & -64.65282 & 0.33$\pm$0.10 & 0.83$\pm$0.48 & 0.40 & 1.00$\pm$0.08 & -0.88$\pm$0.26 & 0.11$\pm$0.02 & 18.42$\pm$0.01 \\
OHSC28 & 88.89700 & -62.34524 & 0.29$\pm$0.01 & 0.38$\pm$0.12 & 0.13 & 2.63$\pm$0.31 & -0.58$\pm$0.01 & 0.05$\pm$0.01 & 18.29$\pm$0.01 \\
BSDL3181 & 88.97086 & -65.29311 & 0.38$\pm$0.06 & 0.73$\pm$0.56 & 0.28 & 0.98$\pm$0.04 & -1.18$\pm$0.14 & 0.14$\pm$0.04 & 18.39$\pm$0.02 \\
KMHK1530 & 89.03412 & -63.64278 & 0.17$\pm$0.06 & 1.58$\pm$0.57 & 0.97 & 2.63$\pm$0.06 & -0.88$\pm$0.01 & 0.06$\pm$0.01 & 18.55$\pm$0.01 \\
DES001SC24 & 89.08272 & -64.71755 & 0.17$\pm$0.06 & 1.18$\pm$0.57 & 0.84 & 1.74$\pm$0.44 & -1.18$\pm$0.08 & 0.13$\pm$0.01 & 18.30$\pm$0.04 \\
BSDL3189 & 89.38447 & -65.34330 & 0.15$\pm$0.04 & 1.29$\pm$0.58 & 0.94 & 1.91$\pm$0.06 & -0.58$\pm$0.01 & 0.06$\pm$0.01 & 18.49$\pm$0.01 \\
LW340,KMHK1548 & 89.39715 & -64.99839 & 0.34$\pm$0.06 & 1.19$\pm$0.49 & 0.55 & 1.58$\pm$0.13 & -0.88$\pm$0.17 & 0.05$\pm$0.05 & 18.49$\pm$0.10 \\
SL798,LW344,KMHK1556 & 89.55014 & -63.89450 & 0.22$\pm$0.05 & 1.22$\pm$0.52 & 0.74 & 2.51$\pm$0.28 & -0.88$\pm$0.05 & 0.03$\pm$0.02 & 18.40$\pm$0.10 \\
DES001SC27 & 89.60803 & -64.59479 & 0.10$\pm$0.03 & 1.05$\pm$0.67 & 1.03 & 1.10$\pm$0.16 & -0.70$\pm$0.34 & 0.08$\pm$0.07 & 18.14$\pm$0.14 \\
NGC2162,SL814,LW351,ESO86SC47 & 90.12589 & -63.72173 & 0.91$\pm$0.03 & 1.86$\pm$0.05 & 0.31 & 1.29$\pm$0.03 & -0.88$\pm$0.01 & 0.06$\pm$0.01 & 18.60$\pm$0.01 \\
KMHK1593 & 90.46676 & -64.13292 & 0.14$\pm$0.03 & 2.34$\pm$0.49 & 1.23 & 3.39$\pm$0.45 & -0.88$\pm$0.10 & 0.05$\pm$0.03 & 18.22$\pm$0.01 \\
ESO121SC03,KMHK1591 & 90.51384 & -60.52379 & 0.47$\pm$0.04 & 2.63$\pm$0.09 & 0.75 & 9.77$\pm$0.30 & -1.40$\pm$0.05 & 0.07$\pm$0.01 & 18.37$\pm$0.01 \\
DES001SC28 & 90.82990 & -64.83138 & 0.15$\pm$0.04 & 1.15$\pm$0.64 & 0.89 & 0.79$\pm$0.09 & -0.70$\pm$0.03 & 0.10$\pm$0.02 & 18.77$\pm$0.10 \\
NGC2193,SL839,LW387,ESO86SC57 & 91.57213 & -65.09874 & 0.46$\pm$0.05 & 3.29$\pm$0.26 & 0.85 & 2.40$\pm$0.06 & -0.70$\pm$0.04 & 0.03$\pm$0.01 & 18.36$\pm$0.01 \\\hline
\end{longtable}
\bsp	% typesetting comment
\label{lastpage}
\end{landscape}
}


\begin{thebibliography}{}
\bibitem[\protect\citeauthoryear{Alves \& Nelson}{2000}]{2000ApJ...542..789A} Alves, D.~R., \& Nelson, C.~A.\ 2000, Astrophysical Journal, 542, 789 
\bibitem[\protect\citeauthoryear{Annis}{2013}]{2013AAS...22133505A} Annis, J.~T.\ 2013, American Astronomical Society Meeting Abstracts, 221, \#335.05 
\bibitem[\protect\citeauthoryear{Annunziatella et al.}{2013}]{2013PASP..125...68A} Annunziatella, M., Mercurio, A., Brescia, M., Cavuoti, S., \& Longo, G.\ 2013, The Publications of the Astronomical Society of the Pacific, 125, 68 
\bibitem[\protect\citeauthoryear{Ashman et al.}{1994}]{1994AJ....108.2348A} Ashman, K.~M., Bird, C.~M., \& Zepf, S.~E.\ 1994, Astronomical Journal, 108, 2348 
\bibitem[\protect\citeauthoryear{Balbinot et al.}{2015}]{2015MNRAS.449.1129B} Balbinot, E., Santiago, B.~X., Girardi, L., et al.\ 2015, Monthly Notices of Royal Astronomical Society, 449, 1129 
\bibitem[\protect\citeauthoryear{Balbinot et al.}{2012}]{2012ASPC..461..287B} Balbinot, E., Santiago, B.~X., Girardi, L., et al.\ 2012, Astronomical Data Analysis Software and Systems XXI, 461, 287 
\bibitem[\protect\citeauthoryear{Baumgardt et al.}{2013}]{2013MNRAS.430..676B} Baumgardt, H., Parmentier, G., Anders, P., \& Grebel, E.~K.\ 2013, Monthly Notices of Royal Astronomical Society, 430, 676 
\bibitem[\protect\citeauthoryear{Bekki \& Chiba}{2005}]{2005MNRAS.356..680B} Bekki, K., \& Chiba, M.\ 2005, Monthly Notices of Royal Astronomical Society, 356, 680 
\bibitem[\protect\citeauthoryear{Bertin \& Arnouts}{1996}]{1996A&AS..117..393B} Bertin, E., \& Arnouts, S.\ 1996, Astronomy and Astrophysics, Supplement, 117, 393 
\bibitem[\protect\citeauthoryear{Bertin}{2011}]{2011ASPC..442..435B} Bertin, E.\ 2011, Astronomical Data Analysis Software and Systems XX, 442, 435
\bibitem[\protect\citeauthoryear{Bica et al.}{2008}]{Bica2008} Bica, E., Bonatto, C., Dutra, C.~M., \& Santos, J.~F.~C.\ 2008, Monthly Notices of Royal Astronomical Society, 389, 678 
\bibitem[\protect\citeauthoryear{Bressan et al.}{2012}]{2012MNRAS.427..127B} Bressan, A., Marigo, P., Girardi, L., et al.\ 2012, Monthly Notices of Royal Astronomical Society, 427, 127 
\bibitem[\protect\citeauthoryear{Caldwell \& Coulson}{1986}]{1986MNRAS.218..223C} Caldwell, J.~A.~R., \& Coulson, I.~M.\ 1986, Monthly Notices of the Royal Astronomical Society, 218, 223 
\bibitem[\protect\citeauthoryear{Cardelli et al.}{1989}]{1989ApJ...345..245C} Cardelli, J.~A., Clayton, G.~C., \& Mathis, J.~S.\ 1989, Astrophysical Journal, 345, 245 
\bibitem[\protect\citeauthoryear{Carrera et al.}{2008}]{2008AJ....135..836C} Carrera, R., Gallart, C., Hardy, E., Aparicio, A., \& Zinn, R.\ 2008, Astronomical Journal, 135, 836 
\bibitem[\protect\citeauthoryear{Carrera et al.}{2011}]{2011AJ....142...61C} Carrera, R., Gallart, C., Aparicio, A., \& Hardy, E.\ 2011, Astronomical Journal, 142, 61
\bibitem[\protect\citeauthoryear{Cioni \& the VMC team}{2015}]{2015arXiv150306972C} Cioni, M.-R.~L., \& the VMC team 2015, arXiv:1503.06972 
\bibitem[\protect\citeauthoryear{Da Costa et al.}{1987}]{1987ApJ...321..735D} Da Costa, G.~S., King, C.~R., \& Mould, J.~R.\ 1987, Astrophysical Journal, 321, 735 
\bibitem[\protect\citeauthoryear{Da Costa}{1991}]{1991IAUS..148..183D} Da Costa, G.~S.\ 1991, The Magellanic Clouds, 148, 183 
\bibitem[\protect\citeauthoryear{Desai et al.}{2012}]{2012ApJ...757...83D} Desai, S., Armstrong, R., Mohr, J.~J., et al.\ 2012, Astrophysical Journal, 757, 83 
\bibitem[\protect\citeauthoryear{de Grijs et al.}{2014}]{2014AJ....147..122D} de Grijs, R., Wicker, J.~E., \& Bono, G.\ 2014, Astronomical Journal, 147, 122 
\bibitem[\protect\citeauthoryear{Elson \& Fall}{1988}]{1988AJ.....96.1383E} Elson, R.~A., \& Fall, S.~M.\ 1988, Astronomical Journal, 96, 1383 
\bibitem[\protect\citeauthoryear{Elson et al.}{1997}]{1997MNRAS.289..157E} Elson, R.~A.~W., Gilmore, G.~F., \& Santiago, B.~X.\ 1997, Monthly Notices of Royal Astronomical Society, 289, 157
\bibitem[\protect\citeauthoryear{Fitzpatrick}{1999}]{1999PASP..111...63F} Fitzpatrick, E.~L.\ 1999, The Publications of the Astronomical Society of the Pacific, 111, 63 
\bibitem[\protect\citeauthoryear{Feigelson \& Jogesh Babu}{2012}]{2012msma.book.....F} Feigelson, E.~D., \& Jogesh Babu, G.\ 2012, Modern Statistical Methods for Astronomy, by Eric D.~Feigelson, G.~Jogesh Babu, Cambridge, UK: Cambridge University Press, 2012
\bibitem[\protect\citeauthoryear{Fisher}{1956}]{F1956} Fisher, R.~A.\ 1956, Statistical Methods and Scientific Inference (Edinburgh: Oliver and Boyd)
\bibitem[\protect\citeauthoryear{Flaugher}{2006}]{2006SPIE.6269E..2CF} Flaugher, B.\ 2006, Society of Photo-Optical Instrumentation Engineers (SPIE) Conference Series, 6269, 62692C 
\bibitem[\protect\citeauthoryear{Flaugher et al.}{2015}]{2015arXiv150402900F} Flaugher, B., Diehl, H.~T., Honscheid, K., et al.\ 2015, arXiv:1504.02900 
\bibitem[\protect\citeauthoryear{Geisler et al.}{1997}]{1997AJ....114.1920G} Geisler, D., Bica, E., Dottori, H., et al.\ 1997, Astronomical Journal, 114, 1920
\bibitem[\protect\citeauthoryear{Geisler et al.}{2007}]{2007IAUS..235...92G} Geisler, D., Grocholski, A.~J., Sarajedini, A., Cole, A.~A., \& Smith, V.~V.\ 2007, IAU Symposium, 235, 92 
\bibitem[\protect\citeauthoryear{Girardi et al.}{1995}]{1995A&A...298...87G} Girardi, L., Chiosi, C., Bertelli, G., \& Bressan, A.\ 1995, Astronomy and Astrophysics, 298, 87 
\bibitem[\protect\citeauthoryear{Girardi et al.}{2002}]{2002A&A...391..195G} Girardi, L., Bertelli, G., Bressan, A., et al.\ 2002, Astronomy and Astrophysics, 391, 195 
\bibitem[\protect\citeauthoryear{Glatt et al.}{2010}]{2010A&A...517A..50G} Glatt, K., Grebel, E.~K., \& Koch, A.\ 2010, Astronomy and Astrophysics, 517, AA50 
\bibitem[\protect\citeauthoryear{Gordon et al.}{2003}]{2003ApJ...594..279G} Gordon, K.~D., Clayton, G.~C., Misselt, K.~A., Landolt, A.~U., \& Wolff, M.~J.\ 2003, Astrophysical Journal, 594, 279 
\bibitem[\protect\citeauthoryear{Grocholski et al.}{2007}]{2007AJ....134..680G} Grocholski, A.~J., Sarajedini, A., Olsen, K.~A.~G., Tiede, G.~P., \& Mancone, C.~L.\ 2007, Astronomical Journal, 134, 680 
\bibitem[\protect\citeauthoryear{Harris \& Zaritsky}{2009}]{2009AJ....138.1243H} Harris, J., \& Zaritsky, D.\ 2009, Astronomical Journal, 138, 1243 
\bibitem[\protect\citeauthoryear{Hill et al.}{1995}]{1995A&A...293..347H} Hill, V., Andrievsky, S., \& Spite, M.\ 1995, Astronomy and Astrophysics, 293, 347
\bibitem[\protect\citeauthoryear{Indu \& Subramaniam}{2011}]{2011A&A...535A.115I} Indu, G., \& Subramaniam, A.\ 2011, Astronomy and Astrophysics, 535, AA115
\bibitem[\protect\citeauthoryear{Kallivayalil et al.}{2013}]{2013ApJ...764..161K} Kallivayalil, N., van der Marel, R.~P., Besla, G., Anderson, J., \& Alcock, C.\ 2013, Astrophysical Journal, 764, 161 
\bibitem[\protect\citeauthoryear{Kharchenko et al.}{2012}]{2012A&A...543A.156K} Kharchenko, N.~V., Piskunov, A.~E., Schilbach, E., R{\"o}ser, S., \& Scholz, R.-D.\ 2012, Astronomy and Astrophysics, 543, A156 
\bibitem[\protect\citeauthoryear{Kerber \& Santiago}{2005}]{2005A&A...435...77K} Kerber, L.~O., \& Santiago, B.~X.\ 2005, Astronomy and Astrophysics, 435, 77 
\bibitem[\protect\citeauthoryear{Kerber \& Santiago}{2006}]{2006A&A...452..155K} Kerber, L.~O., \& Santiago, B.~X.\ 2006, Astronomy and Astrophysics, 452, 155 
\bibitem[\protect\citeauthoryear{Kerber et al.}{2007}]{2007A&A...462..139K} Kerber, L.~O., Santiago, B.~X., \& Brocato, E.\ 2007, Astronomy and Astrophysics, 462, 139 
\bibitem[\protect\citeauthoryear{King}{1962}]{1962AJ.....67..471K} King,~I.\ 1962, Astrophysical Journal, 67, 471 
\bibitem[\protect\citeauthoryear{Klein et al.}{2014}]{2014arXiv1405.1035K} Klein, C.~R., Cenko,  S.~B., Miller, A.~A., Norman, D.~J., \& Bloom, J.~S.\ 2014, arXiv:1405.1035 
\bibitem[\protect\citeauthoryear{Kontizas et al.}{1993}]{1993A&A...269..107K} Kontizas, M., Kontizas, E., \& Michalitsianos, A.~G.\ 1993, Astronomy and Astrophysics, 269, 107 
\bibitem[\protect\citeauthoryear{Kroupa \& Bastian}{1997}]{1997NewA....2...77K} Kroupa, P., \& Bastian, U.\ 1997, New Astronomy, 2, 77
\bibitem[\protect\citeauthoryear{Kroupa}{2001}]{2001MNRAS.322..231K} Kroupa, P.\ 2001, Monthly Notices of Royal Astronomical Society, 322, 231
\bibitem[\protect\citeauthoryear{Kuehn et al.}{2013}]{2013AJ....145..160K} Kuehn, C.~A., Dame, K., Smith, H.~A., et al.\ 2013, Astronomical Journal, 145, 160
\bibitem[\protect\citeauthoryear{Leonardi \& Rose}{2003}]{2003AJ....126.1811L} Leonardi, A.~J., \& Rose, J.~A.\ 2003, Astronomical Journal, 126, 1811 
\bibitem[\protect\citeauthoryear{Lejeune \& Schaerer}{2001}]{2001A&A...366..538L} Lejeune, T., \& Schaerer, D.\ 2001, Astronomy and Astrophysics, 366, 538 
\bibitem[\protect\citeauthoryear{Li et al.}{2014}]{2014ApJ...784..157L} Li, C., de Grijs, R., \& Deng, L.\ 2014, Astrophysical Journal, 784, 157 
\bibitem[\protect\citeauthoryear{Livanou et al.}{2013}]{2013A&A...554A..16L} Livanou, E., Dapergolas, A., Kontizas, M., et al.\ 2013, Astronomy and Astrophysics, 554, A16 
\bibitem[\protect\citeauthoryear{Lupton}{1993}]{Lupton1993} Lupton, R.\ 1993, Statistics in Theory and Practice (Princeton: Princeton University Press)
\bibitem[\protect\citeauthoryear{Mackey \& Gilmore}{2003}]{2003MNRAS.338...85M} Mackey, A.~D., \& Gilmore, G.~F.\ 2003, Monthly Notices of Royal Astronomical Society, 338, 85
\bibitem[\protect\citeauthoryear{van der Marel}{2001}]{2001AJ....122.1827V} van der Marel, R.~P. 2001, Astronomical Journal, 122, 1827
\bibitem[\protect\citeauthoryear{van der Marel et al.}{2002}]{2002AJ....124.2639V} van der Marel, R.~P., Alves, D.~R., Hardy, E., \& Suntzeff, N.~B.\ 2002, Astronomical Journal, 124, 2639 
\bibitem[\protect\citeauthoryear{Marigo et al.}{2008}]{2008A&A...482..883M} Marigo, P., Girardi, L., Bressan, A., et al.\ 2008, Astronomy and Astrophysics, 482, 883 
\bibitem[\protect\citeauthoryear{Martin et al.}{2008}]{2008ApJ...684.1075M} Martin, N.~F., de Jong, J.~T.~A., \& Rix, H.-W.\ 2008, Astrophysical Journal, 684, 1075 
\bibitem[\protect\citeauthoryear{Meschin et al.}{2014}]{2014MNRAS.438.1067M} Meschin, I., Gallart, C., Aparicio, A., et al.\ 2014, Monthly Notices of Royal Astronomical Society, 438, 1067
\bibitem[\protect\citeauthoryear{Milone et al.}{2009}]{2009A&A...497..755M} Milone, A.~P., Bedin, L.~R., Piotto, G., \& Anderson, J.\ 2009, Astronomy and Astrophysics, 497, 755
\bibitem[\protect\citeauthoryear{Miocchi et al.}{2013}]{2013ApJ...774..151M} Miocchi, P., Lanzoni, B., Ferraro, F.~R., et al.\ 2013, Astrophysical Journal, 774, 151 
\bibitem[\protect\citeauthoryear{Mohr et al.}{2012}]{2012SPIE.8451E..0DM} Mohr, J.~J., Armstrong, R., Bertin, E., et al.\ 2012, Society of Photo-Optical Instrumentation Engineers (SPIE) Conference Series, 8451, 84510D 
\bibitem[\protect\citeauthoryear{Nelder \& Mead}{1965}]{NM1965}, Nelder, J. A. \& Mead, R., Computer Journal 7 , 308-313 . 
\bibitem[\protect\citeauthoryear{Nemec \& Nemec}{1991}]{1991PASP..103...95N} Nemec, J., \& Nemec, A.~F.~L.\ 1991, The Publications of the Astronomical Society of the Pacific, 103, 95 
\bibitem[\protect\citeauthoryear{Nidever \& Smash Team}{2015}]{2015ASPC..491..325N} Nidever, D., \& Smash Team 2015, Astronomical Society of the Pacific Conference Series, 491, 325 
\bibitem[\protect\citeauthoryear{Nikolaev et al.}{2004}]{2004ApJ...601..260N} Nikolaev, S., Drake, A.~J., Keller, S.~C., et al.\ 2004, Astrophysical Journal, 601, 260
\bibitem[\protect\citeauthoryear{Olsen \& Salyk}{2002}]{2002AJ....124.2045O} Olsen, K.~A.~G., \& Salyk, C.\ 2002, Astronomical Journal, 124, 2045 
\bibitem[\protect\citeauthoryear{Olszewski et al.}{1991}]{1991AJ....101..515O} Olszewski, E.~W., Schommer, R.~A., Suntzeff, N.~B., \& Harris, H.~C.\ 1991, Astronomical Journal, 101, 515 
\bibitem[\protect\citeauthoryear{Pagel \& Tautvaisiene}{1998}]{1998MNRAS.299..535P} Pagel, B.~E.~J., \& Tautvaisiene, G.\ 1998, Monthly Notices of Royal Astronomical Society, 299, 535 
\bibitem[\protect\citeauthoryear{Palma et al.}{2015}]{2015MNRAS.450.2122P} Palma, T., Clari{\'a}, J.~J., Geisler, D., Gramajo, L.~V., \& Ahumada, A.~V.\ 2015, Monthly Notices of Royal Astronomical Society, 450, 2122 
\bibitem[\protect\citeauthoryear{Piatti et al.}{2002}]{2002MNRAS.329..556P} Piatti, A.~E., Sarajedini, A., Geisler, D., Bica, E., \& Clari{\'a}, J.~J.\ 2002, Monthly Notices of the Royal Astronomical Society, 329, 556 
\bibitem[\protect\citeauthoryear{Piatti et al.}{2009}]{2009A&A...501..585P} Piatti, A.~E., Geisler, D., Sarajedini, A., \& Gallart, C.\ 2009, Astronomy and Astrophysics, 501, 585 
\bibitem[\protect\citeauthoryear{Piatti et al.}{2013}]{2013BAAA...56..275P} Piatti, A.~E., Keller, S.~C., Mackey, A.~D., \& Da Costa, G.~S.\ 2013, Boletin de la Asociacion Argentina de Astronomia La Plata Argentina, 56, 275 
\bibitem[\protect\citeauthoryear{Piatti \& Geisler}{2013}]{2013AJ....145...17P} Piatti, A.~E., \& Geisler, D.\ 2013, Astronomical Journal, 145, 17 
\bibitem[\protect\citeauthoryear{Piatti et al.}{2014}]{2014MNRAS.444.1425P} Piatti, A.~E., Keller, S.~C., Mackey, A.~D., \& Da Costa, G.~S.\ 2014, Monthly Notices of the Royal Astronomical Society, 444, 1425 
\bibitem[\protect\citeauthoryear{Ripepi et al.}{2004}]{2004CoAst.145...24R} Ripepi, V., Monelli, M., Dall'Ora, M., et al.\ 2004, Communications in Asteroseismology, 145, 24 
\bibitem[\protect\citeauthoryear{Rubele et al.}{2012}]{2012A&A...537A.106R} Rubele, S., Kerber, L., Girardi, L., et al.\ 2012, Astronomy and Astrophysics, 537, AA106 
\bibitem[\protect\citeauthoryear{Rykoff \& DES Cluster Working Group}{2014}]{2014AAS...22314106R} Rykoff, E.~S., \& DES Cluster Working Group 2014, American Astronomical Society Meeting Abstracts \#223, 223, \#141.06 
\bibitem[\protect\citeauthoryear{Saha et al.}{2010}]{2010AJ....140.1719S} Saha, A., Olszewski, E.~W., Brondel, B., et al.\ 2010, Astronomical Journal, 140, 1719
\bibitem[\protect\citeauthoryear{Santiago et al.}{2002}]{2002MNRAS.336..139S} Santiago, B., Kerber,  L., Castro, R., \& de Grijs, R.\ 2002, Monthly Notices of Royal Astronomical Society, 336, 139
\bibitem[\protect\citeauthoryear{Santos et al.}{1999}]{1999AJ....117.2841S} Santos, J.~F.~C., Jr.,  Piatti, A.~E., Clari{\'a}, J.~J., et al.\ 1999, Astronomical Journal, 117, 2841 
\bibitem[\protect\citeauthoryear{Schlegel et al.}{1998}]{1998ApJ...500..525S} Schlegel, D.~J., Finkbeiner, D.~P., \& Davis, M.\ 1998, Astrophysical Journal, 500, 525 
\bibitem[\protect\citeauthoryear{Sharma et al.}{2010}]{2010AJ....139..878S} Sharma, S., Borissova, J., Kurtev, R., Ivanov, V.~D., \& Geisler, D.\ 2010, Astronomical Journal, 139, 878 
\bibitem[\protect\citeauthoryear{Soumagnac et al.}{2013}]{2013arXiv1306.5236S} Soumagnac, M.~T., Abdalla, F.~B., Lahav, O., et al.\ 2013, arXiv:1306.5236 
\bibitem[\protect\citeauthoryear{Sprott}{2000}]{S2000} Sprott, D.~A.\ 2000, Statistical Inference in Science (Ontario: Springer)
\bibitem[\protect\citeauthoryear{Stetson}{1987}]{1987PASP...99..191S} Stetson, P.~B.\ 1987, PASP, 99, 191
\bibitem[\protect\citeauthoryear{Subramanian \& Subramaniam}{2009}]{2009A&A...496..399S} Subramanian, S., \& Subramaniam, A.\ 2009, Astronomy and Astrophysics, 496, 399
\bibitem[\protect\citeauthoryear{Subramanian \& Subramaniam}{2010}]{2010A&A...520A..24S} Subramanian, S., \& Subramaniam, A.\ 2010, Astronomy and Astrophysics, 520, A24 
\bibitem[\protect\citeauthoryear{Valcarce et al.}{2012}]{2012A&A...547A...5V} Valcarce, A.~A.~R., Catelan, M., \& Sweigart, A.~V.\ 2012, Astronomy and Astrophysics, 547, AA5 
\bibitem[\protect\citeauthoryear{van den Bergh}{1991}]{1991ApJ...369....1V} van den Bergh, S.\ 1991, Astrophysical Journal, 369, 1 
\bibitem[\protect\citeauthoryear{The Dark Energy Survey Collaboration}{2005}]{2005astro.ph.10346T} The Dark Energy Survey Collaboration 2005, arXiv:astro-ph/0510346 
\bibitem[\protect\citeauthoryear{Tucker et al.}{2014}]{2014AAS...22325411T} Tucker, D.~L., Allam, S.~S., Annis, J.~T., et al.\ 2014, American Astronomical Society Meeting Abstracts, 223, \#254.11 
\bibitem[\protect\citeauthoryear{Walker}{1992}]{1992AJ....103.1166W} Walker, A.~R.\ 1992, Astronomical Journal, 103, 1166
\bibitem[\protect\citeauthoryear{Weinberg \& Nikolaev}{2001}]{2001ApJ...548..712W} Weinberg, M.~D., \& Nikolaev, S.\ 2001, Astrophysical Journal, 548, 712 
\bibitem[\protect\citeauthoryear{Weisz et al.}{2013}]{2013MNRAS.431..364W} Weisz, D.~R., Dolphin, A.~E., Skillman, E.~D., et al.\ 2013, Monthly Notices of Royal Astronomical Society, 431, 364 
\bibitem[\protect\citeauthoryear{Werchan \& Zaritsky}{2011}]{2011AJ....142...48W} Werchan, F., \& Zaritsky, D.\ 2011, Astronomical Journal, 142, 48 
\end{thebibliography}
\end{document}